\documentclass[amsmath, amssymb, aps, longbibliography, prl, superscriptaddress,
reprint,
nobibnotes, 
]{revtex4-2}

\makeatletter
\def\NAT@spacechar{}
\makeatother

\usepackage{siunitx}
\usepackage[colorlinks=true]{hyperref}
\usepackage{graphicx}

\begin{document}

\definecolor{linkcolor}{rgb}{0,0.4470,0.7410}%
\hypersetup{allcolors=linkcolor}

\title{Generation of Large Vortex-Free Superfluid Helium Nanodroplets}

\newcommand{\TUB}{\affiliation{Institute of Optics and Atomic Physics, Technische Universit\"at Berlin, Hardenbergstra\ss{}e 36, 10623 Berlin, Germany}}
\newcommand{\UHH}{\affiliation{Department of Physics, Universit\"at Hamburg, Luruper Chaussee 149, 22761 Hamburg, Germany}}
\newcommand{\MBI}{\affiliation{Max-Born-Institute for Nonlinear Optics and Short Pulse Spectroscopy, Max-Born-Stra\ss{}e 2A, 12489 Berlin, Germany}}
\newcommand{\HRO}{\affiliation{Institute for Physics, University of Rostock, Albert-Einstein-Stra\ss{}e 23, 18059 Rostock, Germany}}
\newcommand{\USCchem}{\affiliation{Department of Chemistry, University of Southern California, 920 Bloom Walk, Los Angeles, California 90089, USA}}
\newcommand{\ETH}{\affiliation{Laboratory for Solid State Physics, Swiss Federal Institute of Technology in Zurich, John-von-Neumann-Weg 9, 8093 Zurich, Switzerland}}
\newcommand{\EuXFEL}{\affiliation{European XFEL, Holzkoppel 4, 22869 Schenefeld, Germany}}
\newcommand{\AAU}{\affiliation{Department of Chemistry, Aarhus University, Langelandsgade 140, 8000 Aarhus C, Denmark}}
\newcommand{\DESY}{\affiliation{Deutsches Elektronen-Synchrotron DESY, Notkestr. 85, 22607 Hamburg, Germany}}
\newcommand{\PNSENSOR}{\affiliation{PNSensor GmbH, Otto-Hahn-Ring 6, 81739 Munich, Germany}}
\newcommand{\USCphys}{\affiliation{Department of Physics and Astronomy, University of Southern California, 920 Bloom Walk, Los Angeles, California 90089, USA}}

\author{Anatoli~Ulmer}
    \email[Corresponding author: ]{anatoli.ulmer@physik.tu-berlin.de}\TUB\UHH{}
\author{Andrea~Heilrath}\TUB{}\MBI{}
\author{Bj\"orn~Senfftleben}\MBI{}\EuXFEL{}
\author{Sean~M.~O.~O'Connell-Lopez}\USCchem{}
\author{Bj\"orn~Kruse}\HRO{}
\author{Lennart~Seiffert}\HRO{}
\author{Katharina~Kolatzki}\MBI\ETH{}
\author{Bruno~Langbehn}\TUB{}
\author{Andreas~Hoffmann}\MBI{}
\author{Thomas~M.~Baumann}\EuXFEL{}
\author{Rebecca~Boll}\EuXFEL{}
\author{Adam~S.~Chatterley}\AAU{}
\author{Alberto~De~Fanis}\EuXFEL{}
\author{Benjamin~Erk}\DESY{}
\author{Swetha~Erukala}\USCchem{}
\author{Alexandra~J.~Feinberg}\USCchem{}
\author{Thomas~Fennel}\HRO{}
\author{Patrik~Grychtol}\EuXFEL{}
\author{Robert~Hartmann}\PNSENSOR{}
\author{Markus~Ilchen}\EuXFEL\DESY{}
\author{Manuel~Izquierdo}\EuXFEL{}
\author{Bennet~Krebs}\HRO{}
\author{Markus~Kuster}\EuXFEL{}
\author{Tommaso~Mazza}\EuXFEL{}
\author{Jacobo~Montaño}\EuXFEL{}
\author{Georg~Noffz}\TUB{}
\author{Daniel~E.~Rivas}\EuXFEL{}
\author{Dieter~Schlosser}\PNSENSOR{}
\author{Fabian~Seel}\TUB{}
\author{Henrik~Stapelfeldt}\AAU{}
\author{Lothar~Str\"uder}\PNSENSOR{}
\author{Josef~Tiggesb\"aumker}\HRO{}
\author{Hazem~Yousef}\EuXFEL{}
\author{Michael~Zabel}\HRO{}
\author{Pawel~Ziołkowski}\EuXFEL{}
\author{Michael~Meyer}\EuXFEL{}
\author{Yevheniy~Ovcharenko}\EuXFEL{}
\author{Andrey~F.~Vilesov}\USCchem\USCphys{}
\author{Thomas~M\"oller}
    \email[Corresponding author: ]{thomas.moeller@physik.tu-berlin.de}\TUB{}
\author{Daniela~Rupp}
    \email[Corresponding author: ]{ruppda@phys.ethz.ch}\MBI\ETH{}
\author{Rico~Mayro~P.~Tanyag}
    \email[Corresponding author: ]{tanyag@physik.tu-berlin.de}\TUB\AAU{}

\date{\today}

\begin{abstract}
    Superfluid helium nanodroplets are an ideal environment for the formation of metastable, self-organized dopant nanostructures. However, the presence of vortices often hinders their formation. Here, we demonstrate the generation of vortex-free helium nanodroplets and explore the size range in which they can be produced. From x-ray diffraction images of xenon-doped droplets, we identify that single compact structures, assigned to vortex-free aggregation, prevail up to $10^8$ atoms per droplet. This finding builds the basis for exploring the assembly of far-from-equilibrium nanostructures at low temperatures. 
\end{abstract}

\maketitle

The visualization of dopant nanostructures formed inside a helium droplet offers an opportunity to parse the factors driving their self-assembly in a cold, superfluid environment, including the effect of intermolecular forces among the dopant materials and their attraction to quantized vortices, if present. While helium nanodroplets have long been used as a matrix for cooling and preparing dopants for their spectroscopic studies~\cite{Toennies2004, Slenczka2022}, it is only through x-ray coherent diffractive imaging that the visualization of both the droplets and dopant nanostructures has become possible~\cite{Gessner2019, Tanyag2022b}. One testament to this imaging technology is the captured in situ configurations of xenon-traced vortex filaments in submicron-sized superfluid helium droplets~\cite{Gomez2014, Tanyag2015, Jones2016, OConnell2020, Feinberg2021, Feinberg2022}.

Vortices with quantized circulation are a manifestation of superfluidity and play an essential role in the rotational dynamics of both Bose--Einstein condensates and superfluid helium~\cite{Donnelly1991, Dalfovo2001, Fetter2009, Barenghi2016}. Although these vortices are fascinating and intriguing by themselves, their presence dominates nanostructure formation since many dopants are easily attracted to them~\cite{Lehmann2003, Coppens2017, Coppens2019, Tanyag2018, Feinberg2021, Feinberg2022}. On the other hand, in their absence, the dopants may be randomly distributed inside the confined droplet space because of the droplet's superfluid state and the weak interaction between helium and the dopant~\cite{Lewerenz1995, Toennies2004, Loginov2011, Slenczka2022}. Additionally, since any heat associated with the doping process and dopant aggregation are rapidly dissipated through the evaporation of helium atoms, the dopants could form amorphous particles with a fractal-like substructure influenced by intermolecular forces~\cite{Lewerenz1995, Nauta2000, Alves2009, Gordon2012, Boltnev2022}. In small helium nanodroplets containing a few tens of thousands of atoms, examples of spectroscopically-identified nanostructures include: silver forming compact clusters at one or several sites in the droplet~\cite{Loginov2011}, hydrogen cyanide assembling into linear chains~\cite{Nauta1999}, water forming the smallest observed ice nanostructure with only six molecules~\cite{Nauta2000}, weakly-bound magnesium aggregating in a foam structure~\cite{Przystawik2008, Gode2013}, and a core-shell structure of a multi-component doped droplet~\cite{Loginov2013, Loginov2017, Haberfehlner2015}.

To explore and image self-organized structures, especially those that can only be formed in superfluid droplets, the presence of vortices needs to be suppressed. One means of controlling their presence is to generate the droplets by expanding cold helium gas. Here, nanodroplets may be stochastically formed from the initial condensation of helium into small clusters that further grow through collision with other helium clusters at some distance away from the nozzle~\cite{Ratner1999}. This way of droplet generation may preclude the acquisition of angular momentum from shear forces prompted by the co-flowing helium fluid. In this paper, we demonstrate that droplets produced from gas condensation using a conical nozzle are larger and have smaller rotational distortion as compared to those produced using a pinhole nozzle in previous experiments~\cite{Gomez2014, Bernando2017, OConnell2020, Verma2020}. Smaller shape deformations indicate a smaller angular momentum of the droplets, which likely contain few if any vortices. Because of this effect, we could identify two major types of xenon nanostructures, filaments and compact, that are respectively assigned to vortex-induced~\cite{Gomez2012, Latimer2014, Volk2016, Tanyag2015, Jones2016, OConnell2020, Feinberg2021, Feinberg2022}, and vortex-free aggregation. Finally, we map the occurrence of these two structures based on the droplet size and find that droplets smaller than $\SI{\sim200}{\nano\metre}$ in diameter are conducive to imaging far-from-equilibrium nanostructures; opening routes for studying self-assembly in a self-bound superfluid droplet.

X-ray coherent diffractive imaging was performed at the Nano-sized Quantum Systems end station of the European XFEL's Small Quantum Systems scientific instrument~\cite{Kuster2021, Mazza2023}. This imaging technique takes snapshots of the droplet size and shape and, if doped, the structure of the dopant aggregates~\cite{Gessner2019, Tanyag2022b}. The helium droplets were produced using a conical nozzle with a throat diameter of $\SI{150}{\micro\metre}$, a half-opening angle of $\ang{3}$, and a channel length of $\SI{9.3}{\milli\metre}$. The nozzle was attached to a Parker valve and operated at a constant stagnation pressure of $\SI{20}{bar}$, while the nozzle temperature $T_0$ was varied from $\SI{5}{\kelvin}$ to $\SI{14}{\kelvin}$. Variable amounts of xenon were introduced in a gas doping cell installed along the droplet's flight path. At $\SI{\sim 1.05}{\text{meters}}$ from the nozzle exit, the droplets reached the interaction volume where they intersected the x-ray beam that had a photon energy of $\SI{1}{\kilo\eV}$, pulse energies in the range of $\SI{3}{\milli\joule}$ to $\SI{5}{\milli\joule}$, and a focus diameter of $\SI{\sim 1.5}{\micro\metre}$ (FWHM)~\cite{Mazza2023}. The scattered light from a diffraction event was collected up to $\ang{6}$ by an area detector located $\SI{\sim 370}{\milli\metre}$ from the interaction volume~\cite{Kuster2021}. A detailed description of the experiment and other relevant information are given in the Supplementary Material (SM)~\cite{SM}.

Figures~\ref{fig:sizesAndShapes}(a) to \ref{fig:sizesAndShapes}(c) show examples of diffraction images for pure (i.e., undoped) droplets produced at $T_0 = \SI{5.0}{\kelvin}$, $\SI{7.3}{\kelvin}$ and $\SI{10.0}{\kelvin}$, respectively. The size distributions are depicted in Fig.~\ref{fig:sizesAndShapes}(d) where the droplet size $N_\mathrm{He}$ is customarily given as the number of helium atoms and is related to the droplet's radius $R_\mathrm{D}$ through $N_\mathrm{He} = (R_\mathrm{D} / \SI{0.222}{\nano\metre})^3$~\cite{Toennies2004, Slenczka2022}. The nozzle temperatures explored here encompass two droplet production regimes: the fragmentation of liquid helium close to the nozzle throat at $T_0 = \SI{5}{\kelvin}$, and the condensation of helium gas after the nozzle throat at $T_0 \geq \SI{10}{\kelvin}$~\cite{Toennies2004, Slenczka2022}. The size distribution at $T_0 = \SI{5}{\kelvin}$ is bimodal with a broad size range, whereas, that for $T_0 \geq\SI{10}{\kelvin}$ is more narrow with an almost log-normal shape. Under similar stagnation conditions, the droplets produced at the gas condensation regime using the conical nozzle are about two to three orders of magnitude larger than those produced using the $\SI{5}{\micro\metre}$ pinhole nozzle~\cite{Toennies2004, Slenczka2022}, also see Fig. S7 in SM~\cite{SM}.


\begin{figure}[htbp]
  \includegraphics[width=8.6cm]{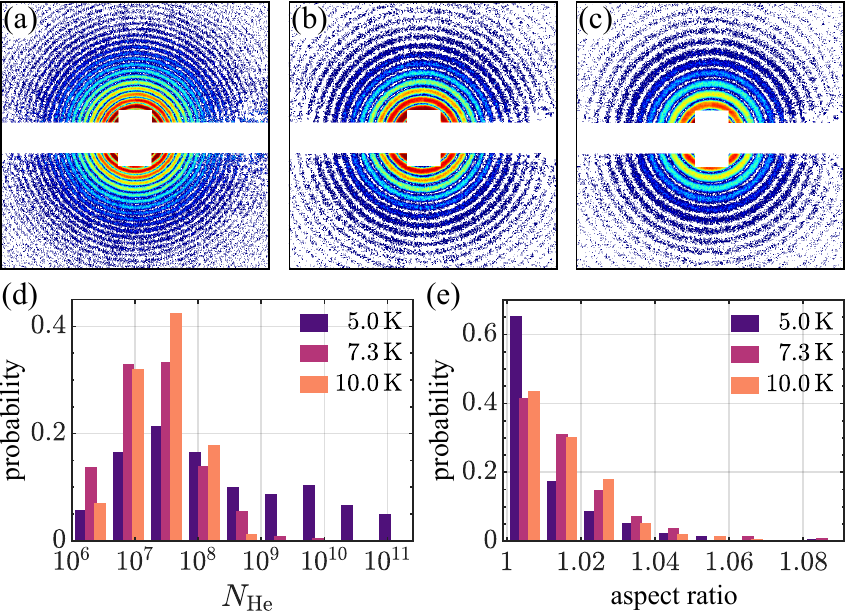}
  \caption{Selected diffraction images of pure droplets obtained at a nozzle temperature of (a) $\SI{5.0}{\kelvin}$, (b) $\SI{7.3}{\kelvin}$, and (c) $\SI{10.0}{\kelvin}$. The size distributions are given in panel (d), and their aspect ratio distributions in panel (e). The numbers of diffraction images used for the size (aspect ratio) distributions are $291$ ($236$) for $\SI{5.0}{\kelvin}$, $766$ ($488$) for $\SI{7.3}{\kelvin}$, and $272$ ($214$) for $\SI{10.0}{\kelvin}$.\label{fig:sizesAndShapes}}
\end{figure}


As with rotating viscous liquid drops~\cite{Brown1980}, a rotating superfluid helium droplet can have a series of equilibrium shapes depending on its angular velocity and angular momentum~\cite{Ancilotto2018, Pi2021}. One way to describe a droplet's shape is through its aspect ratio $AR$, which is defined as the ratio between the semi-major and semi-minor axes of the projected droplet density onto a two-dimensional plane. Figure~\ref{fig:sizesAndShapes}(e) shows the aspect ratio distribution for $T_0 = \SI{5.0}{\kelvin}$, $\SI{7.3}{\kelvin}$ and $\SI{10.0}{\kelvin}$ with average values of $(1.011 \, {}^{+}_{-} \, {}^{0.012}_{0.011})$,  $(1.016 \pm 0.013)$, and $(1.015 \pm 0.012)$, respectively. Collectively, $\num{96.4}\%$ ($\num{99.9}\%$) of the droplets generated using the conical nozzle have an aspect ratio smaller than $\num{1.05}$ ($\num{1.10}$), making them close to spherical, and the largest observed $AR$ is $\num{1.3}$. These values are considerably smaller than those obtained from previous investigations using different kinds of nozzles. For instance, values as high as $\num{2.3}$ have been reported for droplets produced from a $\SI{5}{\micro\metre}$ pinhole nozzle at $\SI{20}{bar}$ and $\SI{5}{\kelvin}$ with $\num{\sim 65}\%$ of the droplets having $AR \leq 1.05$~\cite{Tanyag2018b}. Another experiment using a flow-cryostat but with the same $\SI{5}{\micro\metre}$ pinhole nozzle at $T_0 \simeq \SI{5}{\kelvin}$ reported a mean value $\langle AR \rangle = 1.059 \pm 0.005$, with $1\%$ of the droplets exhibiting $AR \geq 1.4$~\cite{Verma2020}. For droplets produced at $\SI{80}{bar}$ and $\SI{5.4}{\kelvin}$ using a trumpet-shaped nozzle with a throat diameter of $\SI{100}{\micro\metre}$ and a half-opening angle of $\ang{20}$, aspect ratios as large as $\num{\sim 3}$ were observed, and $92.9\%$ of the droplets have close-to-spherical shape~\cite{Langbehn2018}. Further elaboration about the droplets produced using these three nozzles is included in the SM~\cite{SM}.


\begin{figure}[htbp]
  \centering
   \includegraphics[width=8.6cm]{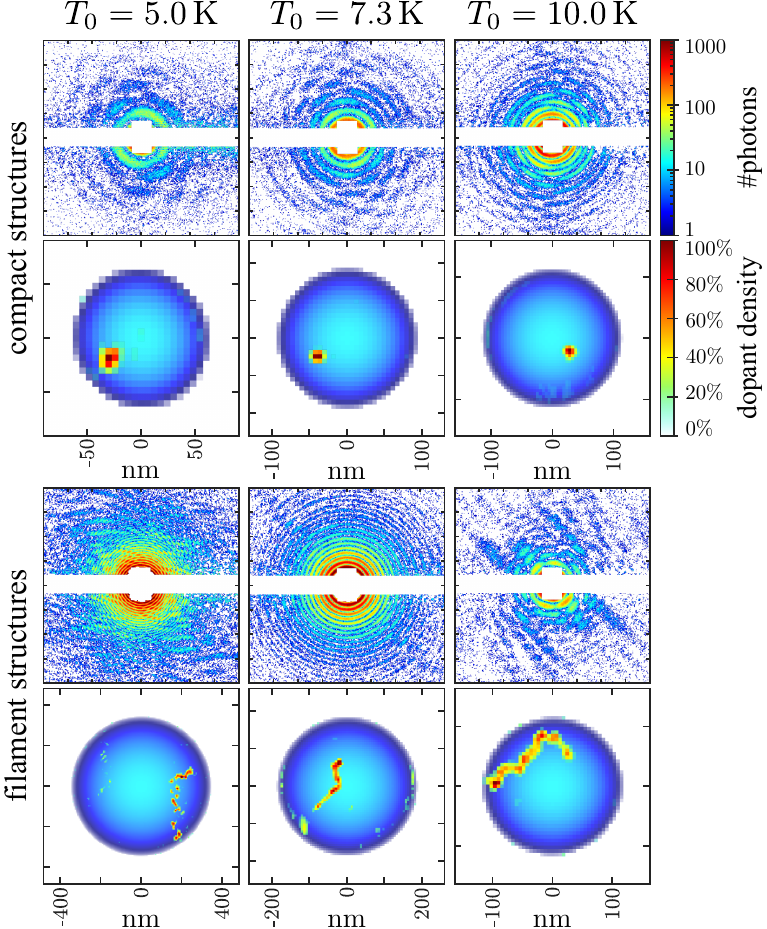}
   \caption{Major types of xenon nanostructures inside helium droplets generated using a conical nozzle at different nozzle temperatures $T_0$. Shown are selected examples of compact structures from vortex-free droplets and filaments from vortex-hosting ones. For each structure type, the first row shows the measured diffraction image, whereas the second shows the numerically reconstructed xenon-doped droplets using a modified DCDI, see the SM~\cite{SM}. The $\si{\nano\metre}$-scale is different in each figure. The droplet is shown in blue colors, while the droplet density scales with the brightness. The color scales for the diffraction and the dopant reconstruction are shown on the upper right portion.\label{fig:dopantStructures}}
\end{figure}


 The consequence of generating close-to-spherical and large superfluid nanodroplets is reflected on the types of nanostructures observed inside them. Figure~\ref{fig:dopantStructures} shows diffraction images and numerically reconstructed xenon clusters in droplets generated at different nozzle temperatures. The image reconstruction algorithm is a modified version of the Droplet Coherent Diffractive Imaging (DCDI)~\cite{Tanyag2015, SM}. In each reconstruction, the droplet is represented by dark blue to blue colors, while the xenon clusters are represented by dark red to light green colors. Two major kinds of structures were observed: compact (upper panel) and filaments (lower panel). The compact structures are assigned to vortex-free aggregation, are located at some distance away from the droplet's center, and have roughly circular shapes with radii ranging from $\SI{10}{\nano\metre}$ to $\SI{15}{\nano\metre}$. Because of the high thermal conductivity of superfluid helium, these compact structures possibly grow with an amorphous or even fractal morphology as similarly observed in bulk superfluid helium~\cite{Gordon2012, vanSciver2009, Boltnev2022}. Compact structures larger than $\SI{20}{\nano\metre}$ are also observed (see Fig. S5 in SM~\cite{SM}), but instead of having a more circular shape, they seem to be composed of two clusters, where each may have initially been formed at different aggregation sites before colliding and fusing into a cluster-cluster aggregate. However, because the heat of condensation is not enough to melt them, the two clusters remain distinct without reforming into one big clump~\cite{Gordon2012}.

 Similar to previous observations~\cite{Gomez2014, Tanyag2015, Jones2016, OConnell2020, Feinberg2021, Feinberg2022}, the filaments are also assigned to xenon clusters tracing the vortex length. In contrast, however, the number of vortices in our experiment is rather few ($<6$), and no Bragg peaks, whose occurrence in a diffraction attests to the presence of a vortex lattice~\cite{Gomez2014, OConnell2020}, were observed. The off-centered single vortices in Fig.~\ref{fig:dopantStructures} convey that the xenon distribution along the vortex line is uneven; rather, each is dotted by distinct nanometer-sized ($10$ to $\SI{20}{\nano\metre}$) xenon clusters. Additionally, the filament is not smooth and its curvature is not as expected for an off-centered vortex~\cite{Bauer1995, Lehmann2003}. Instead, it has oscillations along its length that specify underlying dynamics of how dopants approach a vortex~\cite{Coppens2017, Coppens2019, vanSciver2009, Giuriato2020} and indicate the presence of helical Kelvin waves relevant to quantum turbulence and the cascade of excitations in superfluids~\cite{Donnelly1991, Paolett2011, Madeira2020}. This observation of single vortices is instrumental to understanding vortex nucleation and decay in a superfluid helium droplet, as similarly observed in Bose--Einstein condensates~\cite{Raman2001, Rosenbusch2002}.


\begin{figure}[ht!]
  \centering
  \includegraphics[width=8.6cm]{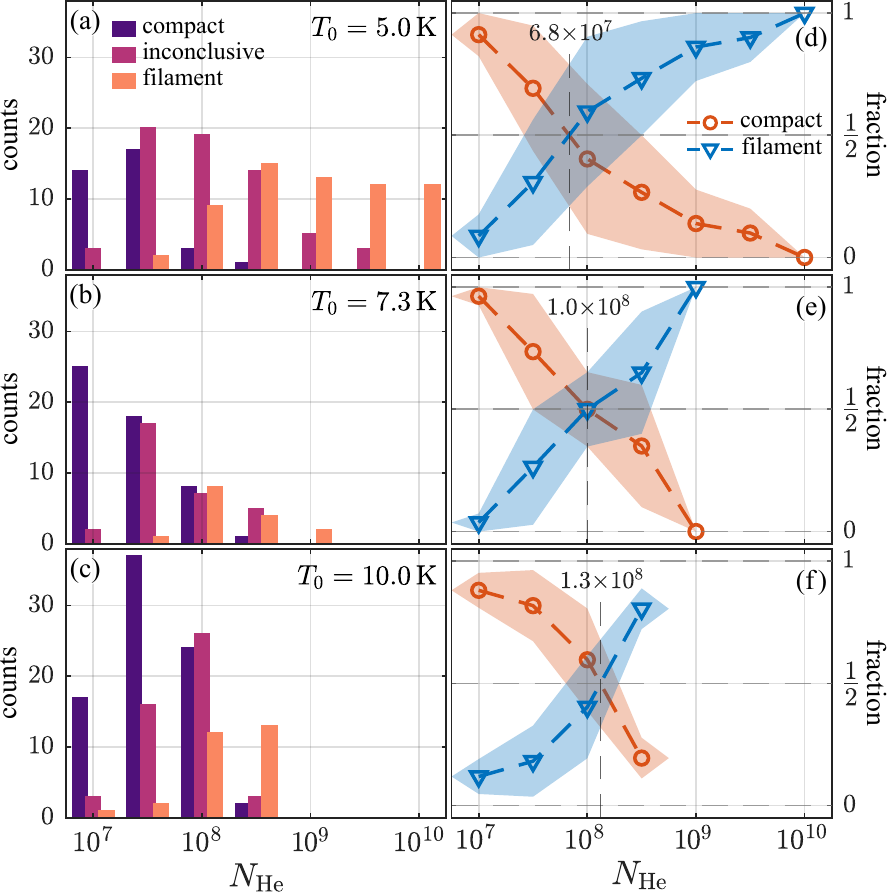}
  \caption{Mapping structure formation as a function of the droplet size. The counts (left) and fractions (right) of the identified compact and filament xenon nanostructures for $T_0 = \SI{5}{\kelvin}$, $\SI{7.3}{\kelvin}$, and $\SI{10}{\kelvin}$ (from top to bottom). Structures that could not be assigned to either categories are categorized as inconclusive (see Fig. S5 in SM~\cite{SM}) and serve as identification error for the fraction diagrams.
  \label{fig:crossing}}
\end{figure}


Crucial to distinguishing these two nanostructures is the droplet's angular momentum, which may be partitioned between quantum vortices and capillary waves. Almost all droplets in earlier imaging experiments were generated in the liquid fragmentation regime and contained multiple vortices~\cite{Gomez2014, Tanyag2015, Jones2016, Bernando2017, Rupp2017, Langbehn2018, OConnell2020, Verma2020, Feinberg2021, Feinberg2022}. These droplets probably acquire angular momentum through the shear flow of the expanding helium fluid~\cite{Gomez2012, Bernando2017, Verma2020}. A smaller nozzle orifice, or higher velocities due to higher stagnation pressures of the expanding fluid, increases the action of shear forces that translates to a larger vorticity of the fluid flow. Within a few nanoseconds during flight, the droplets become superfluid via evaporative cooling and reach a temperature of $\SI{\sim 0.4}{\kelvin}$ after $\SI{\sim 100}{\micro\second}$~\cite{Tanyag2018} with the angular momentum conserved. If the acquired angular momentum is sufficient, one or more stable quantized vortices are nucleated. If more than enough is acquired, multiple vortices form a triangular lattice with the same quantum of circulation and sense of rotation~\cite{Donnelly1991, Barenghi2016} and bear most of the droplet's angular momentum~\cite{Ancilotto2018, OConnell2020, Pi2021}. At high angular momentum, the velocity fields from the vortices and the capillary waves induce deformations that resemble the equilibrium shapes of rotating classical viscous droplets~\cite{Ancilotto2018, OConnell2020, Pi2021}, i.e., from spherical to oblate and to triaxial prolate shapes at higher angular momenta~\cite{Gomez2014, Bernando2017, Langbehn2018, OConnell2020, Verma2020}. The angular momentum required for a single straight vortex to reside in the droplet's center is equal to $N_\mathrm{He}\hbar$~\cite{Seidel1994, Bauer1995, Lehmann2003, Ancilotto2018, Pi2021}. If less angular momentum is present, an off-axis curved vortex may be created that precesses around the droplet's center and whose angular momentum decreases with its curvature~\cite{Bauer1995,Lehmann2003,Coppens2017}. On the other hand, if the acquired angular momentum is insufficient for vortex nucleation, it may be stored as capillary waves, and, in this case, the droplets can only rotate by adopting prolate-shape configurations~\cite{Seidel1994, Ancilotto2018, Pi2021}. In the present experiment, the close-to-spherical droplets suggest that they contain small angular momenta, consistent with the observation of the absence or a small number of vortices, see SM for further discussion~\cite{SM}. However, the exact partition of the small angular momenta between few vortices and capillary waves requires further theoretical and experimental studies, especially for sub-micron to millimeter-sized superfluid drops~\cite{Childress2017}. 

The finding that compact structures are mainly observed in smaller droplets and filaments in larger ones can locate the droplet sizes where the formation of either structure is favored. Figure~\ref{fig:crossing} shows the abundance and fractions of these xenon structures for $T_0 = \SI{5}{\kelvin}$ (top), $\SI{7.3}{\kelvin}$ (middle), and $\SI{10}{\kelvin}$ (bottom). The vertical dashed lines on the fraction panel indicate that the transition crossover, or where the onset of vortex-induced xenon aggregation is facilitated, occurs at $N_\mathrm{He}\approx 10^8$, almost independent of how the droplets are generated. Nevertheless, the means of droplet generation affects the relative abundance of either structures. For instance, filaments are dominant in the liquid fragmentation regime at $T_0 = \SI{5}{\kelvin}$, and conversely, compact structures in the gas condensation regime at $T_0 \geq \SI{10}{\kelvin}$. The observation of filaments in the gas condensation regime is interesting since it suggests that angular momentum was acquired during droplet growth, possibly from droplet collision. A simple kinematic model to estimate the acquired angular momentum shows that for the same collision velocity the angular momentum per atom increases with droplet size, and plausibly why smaller droplets are vortex free~\cite{SM}. In other words, larger droplets require slower collision velocities to have the same angular momentum per atom as compared to smaller droplets. Moreover, these larger droplets also need less angular momentum per atom to maintain a stable curved vortex~\cite{Lehmann2003, SM}. These estimates corroborate the observed size-dependent vortex stability in superfluid helium droplets. Finally, at $\SI{12}{\kelvin}$ and $\SI{20}{bar}$, only compact structures were found, but both structures with a similar crossover were observed again when the stagnation pressure was increased to $\SI{60}{bar}$, see Fig. S8 in SM~\cite{SM}.

In summary, we have demonstrated that large superfluid helium nanodroplets generated using a large-diameter conical nozzle are close to spherical where the angular momentum is likely shared between a few vortices and surface capillary waves. Additionally, we distinguished two main types of dopant aggregation, where their relative abundance is droplet size dependent. The size-dependent transition from vortex-free to vortex-induced structures also emphasizes the size dependence of the rotational state of superfluid helium droplets. For instance, xenon-traced vortex filaments are only readily observed in droplets with more than $10^8$ helium atoms. This size not only benchmarks the onset where vortex-induced nanostructure formation starts to be dominant, but it also indicates the droplet size amenable for the preparation of self-organized nanostructures without the influence of quantum vortices. This study could be a beginning for x-ray imaging of far-from-equilibrium nanostructures from various kinds of dopant materials with different intermolecular forces (e.g. hydrogen or metallic bonding). Without the influence of vortices, it also becomes possible to image dynamics (e.g. nanosplasma ignition) occurring between the dopants and the droplet (in contrast to the observation in Ref.~\cite{Langbehn2022}), and how charges are distributed on the droplet's surface~\cite{Feinberg2022}.

\vspace{2mm}
\begin{acknowledgments}
  We acknowledge the European XFEL in Schenefeld, Germany for the beamtime allocation (SQS~2195) at the SQS instrument, and for the user financial support grant. We would also like to thank the staff, especially Steffen Hauf, for their enormous assistance. We thank the precision mechanics workshops at the TU Berlin Physics Department and Max-Born-Institut for their technical support. Additionally, we greatly benefited from discussions about droplet collisions, nucleation of vortices, and rotation of prolate-shaped droplets from Manuel Barranco, Francesco Ancilotto, and Mart\'i Pi. A.U., R.M.P.T., D.R., T.M\"oller., J.T., and B.Krebs acknowledge funding provided by the Bundesministerium für Bildung und Forschung (BMBF) via grant No. 05K16KT3, the BMBF Forschungsschwerpunkt Freie-Elektronen-Laser FSP-302, Deutsche Forschungsgemeinschaft (DFG) Mo 719/13 and Mo 719/14. A.U. and M.M. acknowledge support by the Cluster of Excellence "Advanced Imaging of Matter" of the DFG - EXC 2056 - project 390715994. Additionally, M.M. acknowledges support by the DFG - SFB-925 - project 170620586. A.Heilrath, B.S., K.K., A.Hoffmann, and D.R. acknowledge funding from the Leibniz-Gemeinschaft via grant No. SAW/2017/MBI4, and K.K. and D.R. further acknowledge the Swiss National Science Foundation under the Grant No. 200021E\_193642. S.M.O.O-L., S.E., A.J.F., and A.F.V. were supported by the National Science Foundation (NSF) under grant nos. CHE-1664990 and CHE-2102318, and S.M.O.O-L. and A.F.V. were also supported by NSF grant nos. DMR-1701077 and DMR-2205081. H.S. and R.M.P.T acknowledge support from Villum Fonden through a Villum Investigator Grant No. 25886. T.F. acknowledges support by the DFG via SFB 1477 “light–matter interactions at interfaces” (ID: 441234705) and via the Heisenberg program (ID: 436382461). M. Ilchen acknowledges funding from the Volkswagen Foundation for a Peter-Paul-Ewald Fellowship. B.Kruse gratefully acknowledges funding by the European Social Fund (ESF) and the Ministry of Education, Science and Culture of Mecklenburg-Western Pomerania (Germany) within the project NEISS (Neural Extraction of Information, Structure and Symmetry in Images) under grant no ESF/14-BM-A55-0007/19. 

  Data recorded for the experiment at the European XFEL are available at \url{https://doi.org/10.22003/XFEL.EU-DATA-002195-00}.

\end{acknowledgments}



%

\end{document}


\definecolor{MatlabBlue}{rgb}{0,0.4470,0.7410}%
\definecolor{linkblue}{rgb}{0.1882, 0.2039, 0.5804}%
\hypersetup{allcolors=MatlabBlue}

\title{Supplementary Material for:\\Generation of Large Vortex-Free Superfluid Helium Nanodroplets}

\newcommand{\TUB}{\affiliation{Institute of Optics and Atomic Physics, Technische Universit\"at Berlin, Hardenbergstra\ss{}e 36, 10623 Berlin, Germany}}
\newcommand{\UHH}{\affiliation{Department of Physics, Universit\"at Hamburg, Luruper Chaussee 149, 22761 Hamburg, Germany}}
\newcommand{\MBI}{\affiliation{Max-Born-Institute for Nonlinear Optics and Short Pulse Spectroscopy, Max-Born-Stra\ss{}e 2A, 12489 Berlin, Germany}}
\newcommand{\HRO}{\affiliation{Institute for Physics, University of Rostock, Albert-Einstein-Stra\ss{}e 23, 18059 Rostock, Germany}}
\newcommand{\USCchem}{\affiliation{Department of Chemistry, University of Southern California, 920 Bloom Walk, Los Angeles, California 90089, USA}}
\newcommand{\ETH}{\affiliation{Laboratory for Solid State Physics, Swiss Federal Institute of Technology in Zurich, John-von-Neumann-Weg 9, 8093 Zurich, Switzerland}}
\newcommand{\EuXFEL}{\affiliation{European XFEL, Holzkoppel 4, 22869 Schenefeld, Germany}}
\newcommand{\AAU}{\affiliation{Department of Chemistry, Aarhus University, Langelandsgade 140, 8000 Aarhus C, Denmark}}
\newcommand{\DESY}{\affiliation{Deutsches Elektronen-Synchrotron DESY, Notkestr. 85, 22607 Hamburg, Germany}}
\newcommand{\PNSENSOR}{\affiliation{PNSensor GmbH, Otto-Hahn-Ring 6, 81739 Munich, Germany}}
\newcommand{\USCphys}{\affiliation{Department of Physics and Astronomy, University of Southern California, 920 Bloom Walk, Los Angeles, California 90089, USA}}

\author{Anatoli~Ulmer}
    \email[Corresponding author: ]{anatoli.ulmer@physik.tu-berlin.de}\TUB\UHH{}
\author{Andrea~Heilrath}\TUB{}\MBI{}
\author{Bj\"orn~Senfftleben}\MBI{}\EuXFEL{}
\author{Sean~M.~O.~O'Connell-Lopez}\USCchem{}
\author{Bj\"orn~Kruse}\HRO{}
\author{Lennart~Seiffert}\HRO{}
\author{Katharina~Kolatzki}\MBI\ETH{}
\author{Bruno~Langbehn}\TUB{}
\author{Andreas~Hoffmann}\MBI{}
\author{Thomas~M.~Baumann}\EuXFEL{}
\author{Rebecca~Boll}\EuXFEL{}
\author{Adam~S.~Chatterley}\AAU{}
\author{Alberto~De~Fanis}\EuXFEL{}
\author{Benjamin~Erk}\DESY{}
\author{Swetha~Erukala}\USCchem{}
\author{Alexandra~J.~Feinberg}\USCchem{}
\author{Thomas~Fennel}\HRO{}
\author{Patrik~Grychtol}\EuXFEL{}
\author{Robert~Hartmann}\PNSENSOR{}
\author{Markus~Ilchen}\EuXFEL\DESY{}
\author{Manuel~Izquierdo}\EuXFEL{}
\author{Bennet~Krebs}\HRO{}
\author{Markus~Kuster}\EuXFEL{}
\author{Tommaso~Mazza}\EuXFEL{}
\author{Jacobo~Montaño}\EuXFEL{}
\author{Georg~Noffz}\TUB{}
\author{Daniel~E.~Rivas}\EuXFEL{}
\author{Dieter~Schlosser}\PNSENSOR{}
\author{Fabian~Seel}\TUB{}
\author{Henrik~Stapelfeldt}\AAU{}
\author{Lothar~Str\"uder}\PNSENSOR{}
\author{Josef~Tiggesb\"aumker}\HRO{}
\author{Hazem~Yousef}\EuXFEL{}
\author{Michael~Zabel}\HRO{}
\author{Pawel~Ziołkowski}\EuXFEL{}
\author{Michael~Meyer}\EuXFEL{}
\author{Yevheniy~Ovcharenko}\EuXFEL{}
\author{Andrey~F.~Vilesov}\USCchem\USCphys{}
\author{Thomas~M\"oller}
    \email[Corresponding author: ]{thomas.moeller@physik.tu-berlin.de}\TUB{}
\author{Daniela~Rupp}
    \email[Corresponding author: ]{ruppda@phys.ethz.ch}\MBI\ETH{}
\author{Rico~Mayro~P.~Tanyag}
    \email[Corresponding author: ]{tanyag@physik.tu-berlin.de}\TUB\AAU{}

\date{\today}

\maketitle  


\section{Experimental Setup}

Single-shot x-ray coherent diffractive imaging of xenon nanostructures assembled inside helium droplets was performed at the Small Quantum Systems (SQS) instrument of the European X-ray Free Electron Laser (XFEL). The SQS instrument is comprised of multiple experimental end stations~\cite{Mazza2023}. For our measurements, we used the Nano-sized Quantum System (NQS) station, which was designed to study nonlinear and time-resolved responses of nanometer-scale atomic and molecular clusters after irradiation with ultrashort and ultrabright x-rays pulses with photon energies ranging from $\SI{500}{eV}$ to $\SI{3}{keV}$~\cite{Mazza2023}. Our experimental setup consisted of three parts: the sample source system, the main chamber where the x-ray beam intersected the sample beam and where several detectors were installed, and the quadrupole mass spectrometer (QMS), which was used for the characterization of the sample beam. Large helium nanodroplets in this study were generated using a non-standard approach, deviating from the typical method of utilizing a $\SI{5}{\micro\metre}$ pinhole nozzle at temperatures below $\SI{7}{K}$. Instead, a significantly larger conical nozzle with a temperature of up to $\SI{15}{K}$ was employed. Such a nozzle allows large clusters/droplets to be grown from the gas phase in a much slower process. It is well-established that they grow by collision and coagulation, and they are expected to have smaller angular momentum, see section "Droplet Growth by Collision and Coagulation" for a more comprehensive discussion of our growth model.

\begin{figure}
  \includegraphics[width=.48\textwidth]{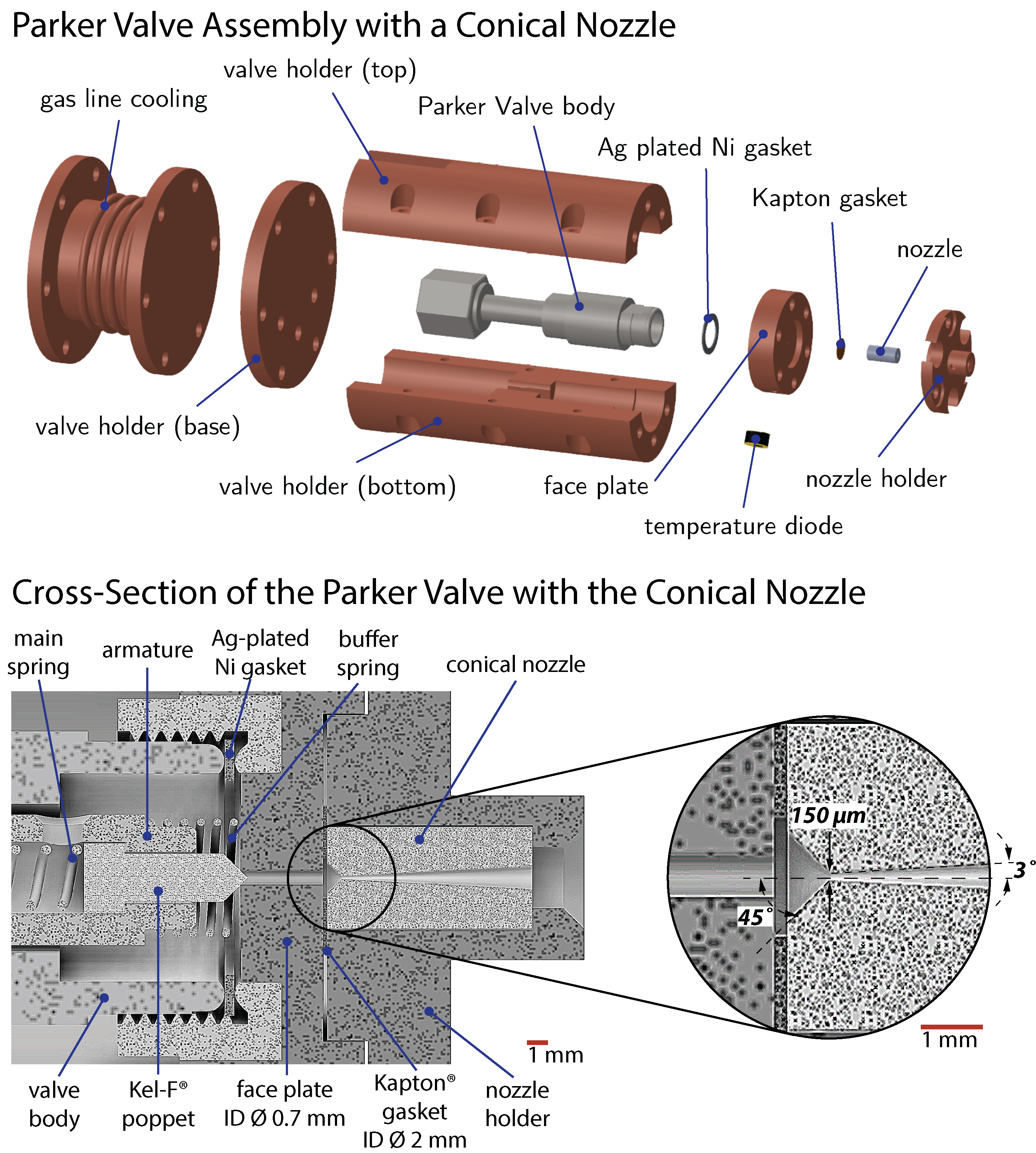}
  \caption{Schematic representation of the Parker Valve assembly with the $\SI{150}{\micro\metre}$ conical nozzle attached. The cross-sectional view displays the closing mechanism of the valve through a Kel-F\textsuperscript{\textregistered{}} poppet, followed by a $\SI{2.2}{\milli\metre}$-long channel with an inner diameter (ID) of $\SI{0.7}{\milli\metre}$. A Kapton\textsuperscript{\textregistered{}} gasket with a $\SI{2}{\milli\metre}$ inner diameter is sandwiched between this channel and the conical nozzle. \label{fig:sm1}}
\end{figure}

The helium droplets were generated by expanding a pre-cooled pressurized gas into the source chamber through a conical nozzle constructed out of stainless steel with a throat diameter of $\SI{150}{\micro\metre}$, a half-opening angle of $\ang{3}$, a channel length of $\SI{9.3}{\milli\metre}$, and an equivalent diameter of $\SI{\sim 2}{\milli\metre}$, calculated using $d_\mathrm{eq} = 0.72 d_\mathrm{throat} / \tan \phi$, where $d_\mathrm{throat}$ and $\phi$ correspond to the nozzle's throat diameter and opening half-angle, respectively~\cite{Hagena1972, Hagena1981, Knuth1997}. This nozzle was perched atop a pulsed valve (Series 99 , Parker Hannifin Corporation), see Fig.~\ref{fig:sm1}. An oxygen-free copper casing enclosed the nozzle and valve assembly and was connected to the second stage of a closed-cycle cryocooler (RDK415E, Sumitomo Heavy Industries). In this experiment, the stagnation pressure was varied from $p_{0}=\SI{20}{bar}$ to $\SI{60}{bar}$, and the nozzle temperature from $T_{0}=\SI{5}{K}$ to $\SI{14}{K}$. An IOTA One pulse driver (Parker Hannifin Corporation) controlled the opening time and cycle of the valve. The driver operation settings were tuned to minimize heat produced by the valve operation. Specifically, about $\SI{290}{V}$ was initially applied to initiate the opening of the valve for $\SI{80}{\micro\second}$, followed by a holding voltage of $\SI{12}{V}$ for another $\SI{560}{\micro\second}$. As determined in exploratory measurements, the total opening time of $\SI{640}{\micro\second}$ should guarantee continuous flow conditions. With these settings and at an opening cycle of $\SI{10}{Hz}$, the valve could be cooled to $ \SI{\sim 7.3}{K}$ at a stagnation pressure of $\SI{20}{bar}$. The nozzle can be cooled further to $\SI{5}{K}$ by reducing the valve's opening cycle to $\SI{2}{Hz}$. For operating the valve at temperatures greater than $\SI{7.3}{K}$, the valve was heated using resistance heating cartridges attached to the nozzle assembly. A comparison of the droplets generated using different nozzle shapes is given at the end of this supplement. 

Once formed, the droplets passed through a $\SI{1}{\milli\metre}$ diameter skimmer, located about $\SI{20}{\milli\metre}$ from the nozzle, into the doping chamber. Here, a $\SI{100}{\milli\metre}$ long, cylindrical gas cell with an entrance and exits holes of $\SI{3}{\milli\metre}$ and $\SI{5}{\milli\metre}$, respectively, was installed. The gas cell was positioned about $\SI{300}{\milli\metre}$ from the nozzle, and its holes are large enough to allow the droplet stream to pass unobstructed. The helium droplets could pick up xenon atoms through inelastic collisions as they pass through the gas cell. Additionally, the doping level was controlled by regulating the xenon pressure in the cell. A differential pumping stage was installed after the doping chamber with a $\SI{4}{\milli\metre}$ skimmer separating them. The differential stage was about $\SI{500}{\milli\metre}$ long, half of which protrudes into the main chamber.

In the main chamber, the helium droplet beam crossed the x-ray pulses. The x-ray photons were generated by the SASE3 undulator of the European XFEL, which for this experiment was operated with a photon energy of $\SI{1}{keV}$, a nominal pulse duration of $\SI{25}{\femto\second}$ (estimated based on the electron bunch charge), and an initial pulse energy of $\SI{3}{\milli\joule}$ to $\SI{5}{\milli\joule}$ as measured after the undulators. A pair of Kirkpatrick-Baez mirrors focuses the x-ray beam to a diameter of $ \SI{\sim 1.5}{\micro\metre}$ with a Rayleigh length of about $\SI{10}{\milli\metre}$. During a diffraction event, a single droplet was hit by a single XFEL pulse, and the scattered photons were collected by a pnCCD (pn-junction Charge Couple Device) detector located $\SI{370}{\milli\metre}$ after the interaction point~\cite{Kuster2021}. The number of x-ray pulses per train was increased to five when the nozzle is operated at $\SI{2}{Hz}$ to increase the probability of a hit. Note that the time delay between the arrival of the x-ray pulse and the different portions of the droplet beam due to the opening of the valve can be varied. Finally, the droplet beam terminated in a diagnostic chamber that housed an Extrel Max 500 QMS, which was used for determining the average droplet size following the droplet titration technique~\cite{Gomez2011}.

\section{Image Data Processing}

\subsection{Raw Data Preprocessing and Hit Finding}

The complete data collected and stored from the pnCCD is a set of hits and non-hits. Diffraction images of pure and doped helium droplets belong to the hit subset, whereas the non-hit subset contains empty frames. Before hits were searched, the readout artifacts from the raw pnCCD image data were first removed by applying pedestal, common mode, and gain corrections. Additionally, the intensities were converted to photon counts. More details on the preprocessing of the pnCCD images are described in Ref.~\cite{Kuster2021}.

After image preprocessing, hits were identified by the number of lit pixels, i.e., number of pixels with values larger than $0.5$ photons. A threshold of $4000$ lit pixels usually yields sufficient signal to determine, at least, the droplet size from a diffraction image. The background stray light for each diffraction event was calculated from the average of the 50 nearest non-hit events and removed using a discrete photon countdown method. In this method, the background and the diffraction image were first divided into chunks of $8 \times 8$ pixels, respectively. The average number of background photons was calculated. In each chunk, a pixel is randomly selected and a photon is subtracted one by one until the number of background photons is eliminated or until no photon is left in the same area.

The hit rate depends on the droplet number density inside the focal volume of the FEL~\cite{Tanyag2018, Tanyag2022b} and is on the order of $0.2 \%$ in our measurements. Such a low hit rate suggests a low droplet number density, making it highly unlikely to observe multiple droplets arriving concurrently in the interaction region.

\begin{figure*}[ht]
  \includegraphics[width=.98\textwidth]{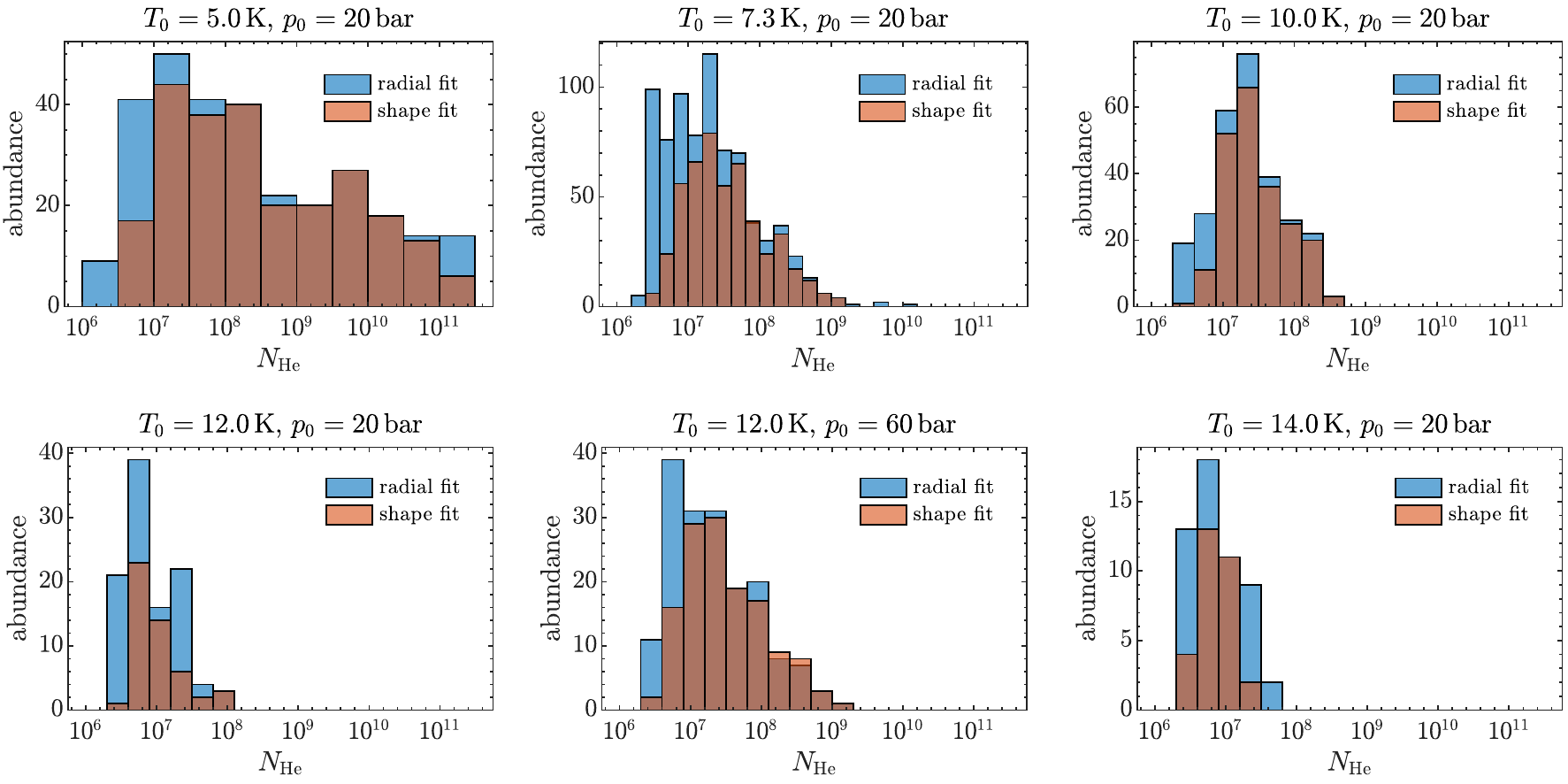}
  \caption{Size distributions of pure helium droplets estimated from radial diffraction profiles (blue) and from the autocorrelation function (red). \label{fig:sm2}}
\end{figure*}

\subsection{Image Centering}

The pnCCD detector in our imaging setup consisted of two autonomous panels, each with $512 \times 1024$ pixels~\cite{Kuster2021}. The relative arrangement between these panels was obtained by considering a subset of diffraction images from a pure droplet and independently fitting ellipses into the diffraction contours lying on each half. The centers of the ellipses were then overlapped, where the relative separation was determined. Finally, this separation was used to recalibrate the absolute encoder positions of the motors controlling the detector and was then applied to all other images within the same data set.

Due to the pointing instability of the x-ray beam, the exact center of a diffraction pattern could vary by $\pm 5$ pixels from shot to shot. Since inaccurate centering of the diffraction leads to spurious image reconstruction, the exact center of each image was determined iteratively. First, a $300 \times 300$ pixels region of interest (ROI) was selected around the initial center position. This ROI included the gap between the two halves of the detector. Then, the upper ROI half $(150 \times 300$ pixels$)$ and the $\ang{180}$ rotated lower half were cross correlated to determine a better diffraction center point. This procedure normally converged quickly to the best center point after a few iterations. For large droplets or those with strong absorption effects inside the droplet, the diffraction patterns can break Friedel's symmetry and can lead to incorrect center determination. In this case, the center position was adjusted manually until the phase ramps, introduced by displacements in the Fourier domain, were eliminated in the spatial domain.

\subsection{Droplet Size and Shape Determination}

Because pure helium droplets are mostly spheroidal with a homogeneous density and negligible absorption, their radial scattering profile can be approximated as that of a non-absorbing sphere, described as~\cite{vandeHulst1957}
\begin{equation}\label{eq:intensityProfile}
  I_{\mathrm{sph}}(q)=A \, V_{\mathrm{sph}}^{2}\left[3\frac{\sin (q R)-(q R) \cos (q R)}{(q R)^{3}}\right]^{2},
\end{equation}
where $R$ is the droplet's radius and $q=4 \pi \lambda^{-1} \sin (\theta / 2)$ is the scattering vector for wavelength $\lambda$ into angle $\theta$. Here, $V_{\mathrm{sph}}=4 \pi R^{3} / 3$ is the droplet volume and $A$ is an amplitude parameter. The droplet radius was extracted from the diffraction images of pure helium droplet by fitting Eq.~\eqref{eq:intensityProfile} into their radial profiles. Figure~\ref{fig:sm2} shows the size distributions of the helium droplets generated using a conical nozzle at six different droplet source conditions used in this experiment; a total of 1640 images was processed for droplet size determination. Again, the droplet size $N_\mathrm{He}$ is given as the number of helium atoms in a droplet and is related to the droplet's radius through $N_\mathrm{He} = (R / \SI{0.222}{\nano\metre})^3$~\cite{Toennies2004, Tanyag2018}. The arithmetic average droplet size $\mean{N_\mathrm{He}}$ are reported and gives an upper limit to the statistics.

\begin{figure}[b]
  \includegraphics[trim=4mm 7mm 12.5mm 8mm, clip, width=.48\textwidth]{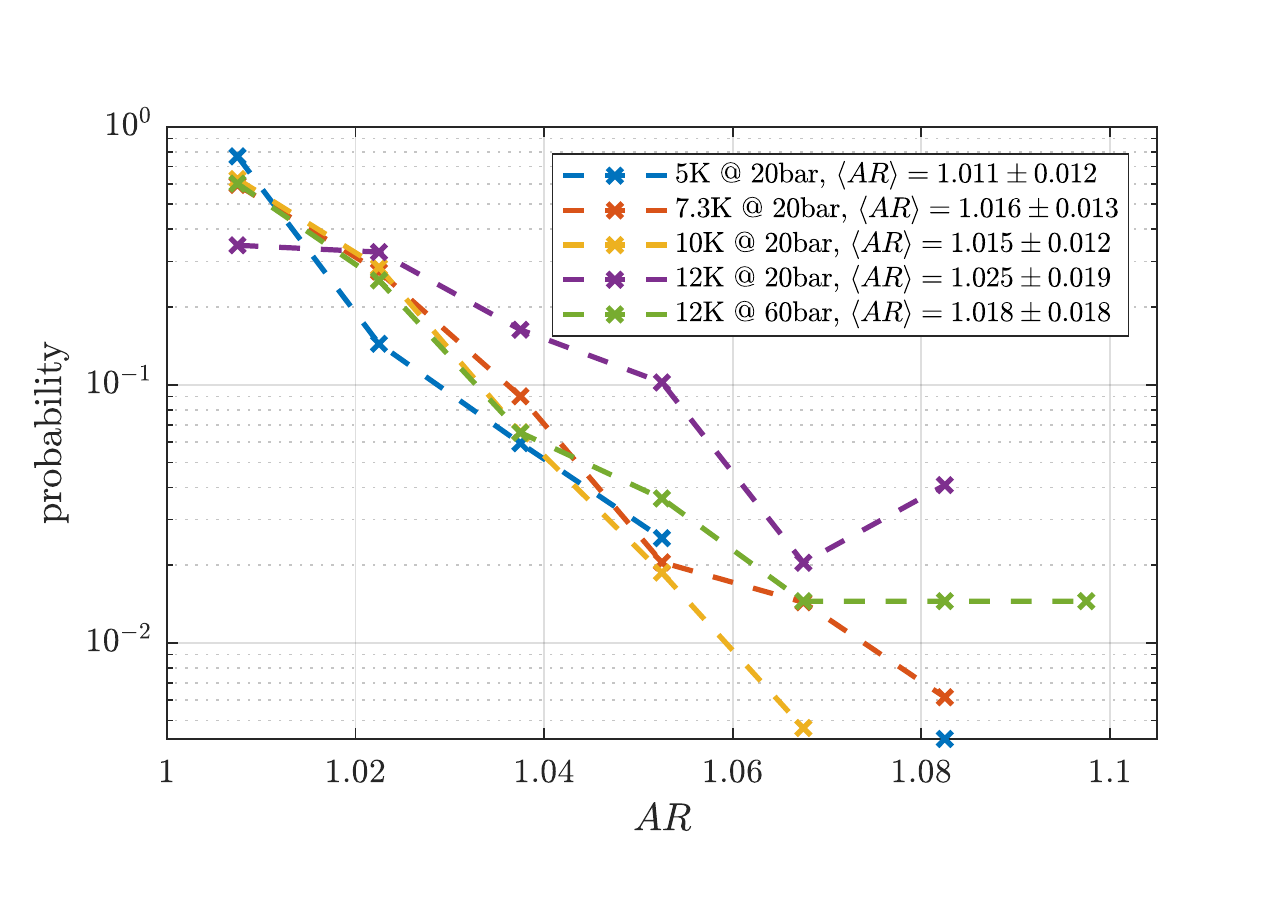}
  \caption{Aspect ratio distributions of the two-dimensional projections of pure droplets in this experiment. The aspect ratio of a droplet was determined from the autocorrelation of its diffraction image. \label{fig:sm3}}
\end{figure}

The shape of a droplet was obtained from the autocorrelation (inverse Fourier transform) of its diffraction pattern, whose extent is twice the size of a droplet's projection onto the detector plane. Because of the central detector hole, the diffraction is high-pass filtered, and its inverse Fourier transform shows an outline. This outline was isolated by applying edge detection techniques and used as input for an elliptical fit that yields the semi-major axis $a$ and semi-minor axis $b$. Data points within $\ang{30}$ of the detector gap axis were excluded in the fitting. A total number of $1155$ images of undoped droplets yielded satisfactory results and were used for determining the droplet's size and shape distributions. A droplet's mean radius was calculated as $R = (a+b)/2$ and the aspect ratio of the projection onto the detector plane as $AR = a / b$. The resulting size distributions are compared to the distributions from radial diffraction profiles in Fig.~\ref{fig:sm2}. Since shape determination from autocorrelation is more susceptible to noise, the sample size is smaller, especially for smaller droplets with $N_{\mathrm{He}} < 10^{7}$. Nevertheless, both distributions show the same overall behavior. The aspect ratio distributions of the undoped droplets in this experiment are shown in Fig.~\ref{fig:sm3}.

\subsection{Image Reconstruction for Doped Droplets}

\begin{figure}
  \includegraphics[width=.35\textwidth]{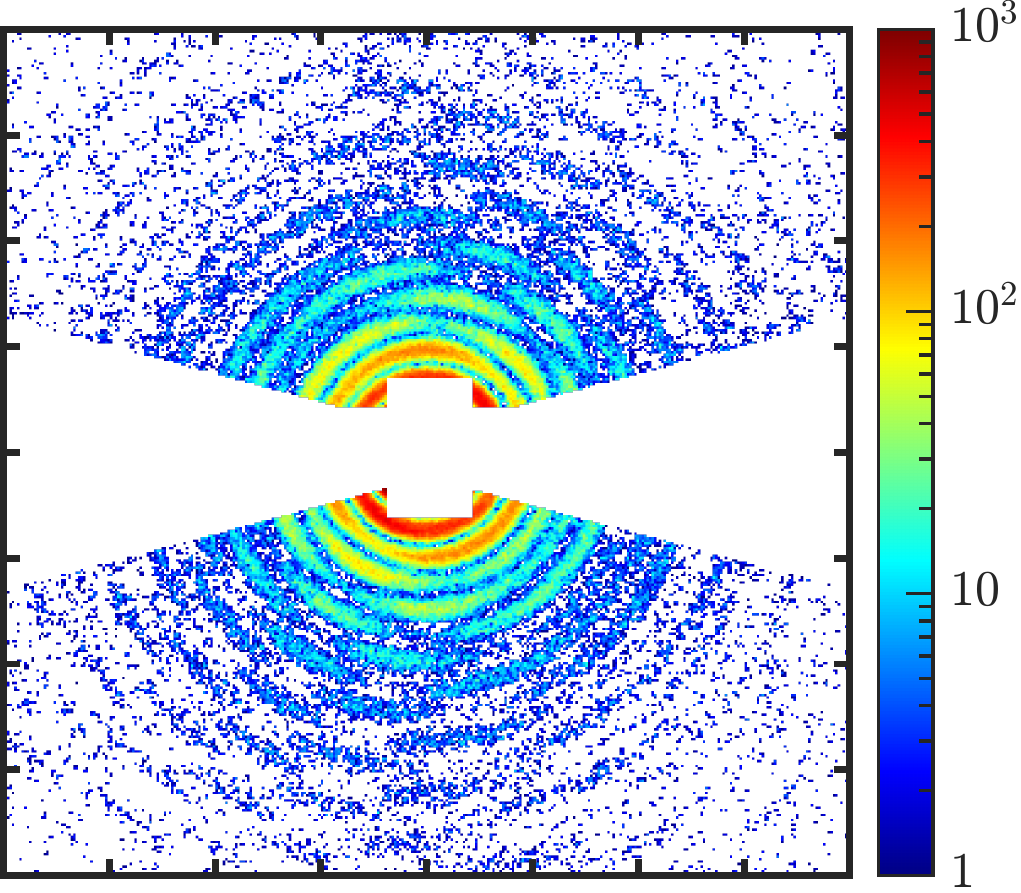}%
  \includegraphics[width=.13\textwidth]{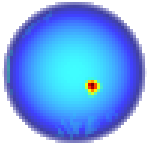}
  \caption{Diffraction image of a xenon-doped droplet with a bow tie-shaped mask (left), and the 
  corresponding DCDI reconstruction (right). The droplet is shown in blue and hosts a compact 
  xenon nanocluster shown in a gradient of colors from green to red. \label{fig:sm4}}
\end{figure}

Prior to numerical image reconstruction, the sizes and shapes of 746 xenon-doped droplets were determined using the autocorrelation method described above. Each diffraction image was assembled, centered, and cropped to an area of $1024 \times 1024$ pixels before rebinning to $512 \times 512$ pixels to improve the signal-to-noise ratio. Then, a bow tie-shaped mask was applied to mask out areas corresponding to the detector hole, the gap between the pnCCD panels, and the regions with strong background stray light, see example in Fig.~\ref{fig:sm4}. Additionally, a Gaussian low-pass filter with $\sigma_{\mathrm{LP}}=40\,$ pixels was applied to the outer region of the diffraction pattern, retaining only diffraction angles $\theta \leq \ang{5}$.

\begin{figure}
    \centering
    \includegraphics[width=8.6cm]{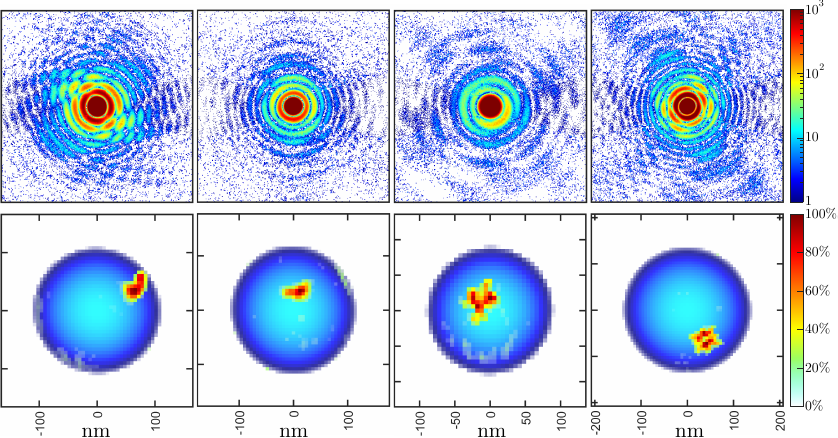}
    \caption{Diffraction images and DCDI reconstructions of doped droplets that were classified as inconclusive, since they could not be identified as either 'compact' or 'filament' structure.}
    \label{fig:sm5}
\end{figure}

The two main categories of dopant structures, as displayed in Fig.~2 of the main text, are: compact xenon clusters, which are associated with vortex-free aggregation; and xenon-traced vortex filaments, which are associated with vortex-induced aggregation. To ensure that vortices are not incorrectly categorized as compact clusters due to poor resolution, diffraction images with low signal-to-noise ratios were filtered out. The achievable intensity limited resolution $d_{\mathrm{SNR}}=$ $2 \pi / q_{\mathrm{SNR}}$ depends on the largest momentum transfer $q_{\mathrm{SNR}}$ (i.e., diffraction angle) with sufficient intensity. Assuming that statistical noise has the biggest contribution, signal-to-noise ratio can be estimated from the radial diffraction profile $I(q)$. Since the relevant quantity for iterative phase reconstructions is the modulus of the diffraction amplitude instead of the intensity, the Poisson signal-to-noise ratio is given by $\operatorname{SNR}\left(\sqrt{P\left(q_{\mathrm{SNR}}\right)}\right)=  2 \sqrt{P\left(q_{\mathrm{SNR}}\right)}$, with $P(q)$ being the number of photons per pixel~\cite{Starodub2008}. Thus, the lower resolution limit can be estimated by choosing an intensity-dependent threshold, e.g., the Rose criterion $\operatorname{SNR}(\mathrm{q}) \geq 5$~\cite{Rose1948}, and deriving $q_{\mathrm{SNR}}$ from it. Diffraction images with insufficient resolution for unambiguous categorization as well as cases where the droplet size was in the order of the resolution limit were excluded in this step, leaving $392$ scattering patterns whose reconstructed doping structures are well above noise. Resolutions with much smaller signal-to-noise ratios have been reported previously (e.g., $1$ photon/pixel in~\cite{Chapman2006}) and further down-binning to the Shannon pixel would also increase the signal-to-noise ratio, leading to considerably higher resolutions than that estimated with this approach; see discussion in Refs.~\cite{Schropp2010} and~\cite{Starodub2008}. From the 392 reconstructions, 330 could be classified in either one of the dopant categories; their corresponding counts are shown in Fig.~3 in the main text. Structures that cannot be conclusively classified as either compact or filament are shown in Fig.~\ref{fig:sm5} and serve as a classification error.

\begin{table*}
  \begin{center}
          \caption{The arithmetic average droplet size $\mean{N_\mathrm{He}}$, ensemble aspect ratio $\mean{AR}$, average aspect ratio $\mean{ar}$, reduced angular momentum $\Lambda_\mathrm{RBR}$ obtained from rigid-body rotation RBR of viscous droplets~\cite{Brown1980}, estimated rigid-body angular momentum $L_\mathrm{RBR}$ per atom, reduced angular momentum $\Lambda_\mathrm{CAP}$ associated with capillary waves in nonaxisymmetric, vortex-free superfluid droplets~\cite{Ancilotto2018, Pi2021}, and estimated angular momentum $L_\mathrm{CAP}$ per atom of a vortex-free droplet are listed for different source stagnation conditions.\label{tab:sm1}}
\begin{tabularx}{\textwidth}{@{\extracolsep{\fill}}ccccccccc}
	\toprule 
	$T_0 / \si{\kelvin}$ & $p_0 / \si{bar}$ & $\mean{N_\mathrm{He}}$ & $\mean{AR}$ & $\mean{ar}$ & $\Lambda_\mathrm{RBR}$ & $L_\mathrm{RBR}/\hbar N_\mathrm{He}$ & $\Lambda_\mathrm{CAP}$ & $L_\mathrm{CAP}/\hbar N_\mathrm{He}$ \\ 
	\midrule 
 
	$5.0$  & $20$ & $\num{7.2e9}$ & $1.011$ & $1.017$ & $0.22$ & $9.5$ & $\num{3.5e-3}$ & $0.15$ \\ 
	$7.3$  & $20$ & $\num{8.8e7}$ & $1.016$ & $1.024$ & $0.26$ & $5.4$ & $\num{4.6e-3}$ & $0.10$ \\ 
	$10.0$ & $20$ & $\num{5.0e7}$ & $1.015$ & $1.022$ & $0.25$ & $4.7$ & $\num{4.2e-3}$ & $0.08$ \\ 
	$12.0$ & $20$ & $\num{1.6e7}$ & $1.027$ & $1.040$ & $0.34$ & $5.3$ & $\num{7.1e-3}$ & $0.11$ \\ 
	$12.0$ & $60$ & $\num{8.0e7}$ & $1.018$ & $1.027$ & $0.27$ & $5.6$ & $\num{5.0e-3}$ & $0.10$ \\ 

	\bottomrule 
\end{tabularx}
\end{center}
\end{table*}

The numerical image reconstructions were calculated using a modified version of the droplet coherent diffractive imaging algorithm (DCDI)~\cite{Tanyag2015}. The calculated projection of the droplet density $\rho_{\mathrm{He}}$ from the shape fit described above was used as a starting condition in the spatial domain. In the frequency domain, the Fourier transform of the droplet density $\tilde{\rho}_{\mathrm{He}}$ was used to obtain the initial phases and for the amplitudes in the masked regions. The density was normalized to the intensity of the diffraction pattern in the unmasked region. Since scattering from dopants contributes to the total intensity, the diffraction from the droplets is overestimated and is corrected by a scaling parameter $C$ to account for the contribution of the dopants to the scattering. For most reconstructions, the value of $C$ was set to $0.9$; in some cases, the value was varied between $0.85$ and $1$. To be consistent with the Rose criterion, the noise threshold parameter was set to $\delta=5 \sigma_{\mathrm{noise}}$, where 
\begin{equation}
  \sigma^2_{\mathrm{noise}} = \frac{1}{N} \sum_{m}^{N} \sigma_{m}^{2}\left(\sqrt{I\left(q_{m}\right)}\right)
\end{equation}
is the standard deviation of the image noise~\cite{Martin2012}, and $\sigma_{m}$ the standard deviation of the diffraction amplitude of pixel $m$; for details, see Ref.~\cite{Martin2012}. Additionally, the positivity constraint was applied in the spatial domain for the real and imaginary, respectively. Finally, the DCDI algorithm was iterated for 100 cycles, which are sufficient for convergence~\cite{Tanyag2015}.

\section{Angular Momentum Estimations}

\subsection{From Droplet Shape}

The droplets generated using a $\SI{150}{\micro\metre}$ conical nozzle have evidently smaller aspect ratios as compared to those generated in previous experiments using a $\SI{5}{\micro\metre}$ pinhole nozzle~\cite{Gomez2014,Bernando2017,Tanyag2018b,Verma2020} at the same stagnation conditions. This already suggests that the means of droplet generation is crucial to controlling the droplet's rotational state. A more deformed droplet generally indicates larger angular momentum due to the presence of multiple vortices, whereas, a close-to-spherical droplet has small angular momentum and may have few or no vortices in addition to the presence of capillary waves on the droplet's surface~\cite{Seidel1994, Ancilotto2018, Pi2021}. To benchmark the rotational state of the droplets produced here, their angular momentum $L$ is estimated from their mean droplet size $\left\langle N_{\mathrm{He}} \right\rangle$ and aspect ratio $\langle AR \rangle$, similar to the approach performed in Refs.~\cite{OConnell2020, Verma2020}. Starting from the average aspect ratio $\langle AR\rangle$ obtained from the two-dimensional projection of the droplet's shape on the detector plane, the three-dimensional average aspect ratio can be estimated using $\langle ar \rangle \approx 1+\frac{3}{2} \times(\langle AR \rangle - 1)$~\cite{Verma2020}, from which the reduced angular momentum $\Lambda$ are determined from tabulated values. Then, the average angular momentum $L$ is calculated using~\cite{Brown1980}
\begin{equation}\label{eq:reducedAngular}
  \Lambda=\frac{L}{\sqrt{8 \sigma_{\mathrm{st}} \rho_{\mathrm{He}} R^{7}}},
\end{equation}
where $\rho_{\mathrm{He}}=\SI{145}{\kilogram \, \metre^{-3}}$ is the bulk liquid helium density at low temperatures, $R = r_{0} N_{\mathrm{He}}^{1/3}$ is the droplet radius with $r_{0} = \left(4 \pi \rho_{\mathrm{He}} / 3 m_{4}\right)^{-1/3}$ and $m_{4}$ being the ${ }^{4} \mathrm{He}$ mass, and $\sigma_{\mathrm{st}} = \SI{3.54e-4}{\newton \, \metre^{-1}}$ is the surface tension of helium at low temperatures~\cite{Donnelly1998}. Two sets of tabulated $\Lambda$ values are used: the first considers a droplet following a rigid-body rotation and thus conforms to the stability curve of a rotating viscous droplet~\cite{Brown1980} (see Figs.~S3(A) and (B) in Ref.~\cite{Gomez2014} and Fig.~4 in Ref.~\cite{Verma2020}), while the second assumes that all angular momenta is carried by capillary waves on the droplet's surface~\cite{Seidel1994, Ancilotto2018, Pi2021}.  The results of these estimates are tabulated in Table~\ref{tab:sm1}. The calculated angular momenta using the rigid-body approach suggest that multiple vortices $(\num{\sim 15})$ should have been observed in the experiment, where $N_\mathrm{vort} \approx 2.5\, L_\mathrm{RBR} / \hbar N_\mathrm{He}$~\cite{OConnell2020}. On the other hand, the estimates from the capillary waves approach indicate that no straight vortex would be expected since it requires an angular momentum per atom of $\SI{1}{\hbar}$~\cite{Seidel1994, Bauer1995, Lehmann2003}. Calculations are available on how the angular momentum is partitioned between the co-existence of vortices and capillary waves in prolate-shaped droplets, and how the presence of multiple vortices accommodates most of the angular momentum of the droplet~\cite{OConnell2020, Pi2021}. However, these calculation were only performed in rather smaller droplets containing a few tens of thousand atoms given the exorbitant computational cost associated in performing DFT calculations and may not be easily extendable for droplet sizes relevant to this experiment (${\gtrsim}10^7$ atoms). Therefore, the angular momentum calculation of these large superfluid droplets, which may contain a combination of few vortices and surface capillary waves requires further study, considering especially that the computational rubric must consider the interplay between the position of an off-centered curved vortex in the droplet and the droplet shape itself.

\subsection{From Droplet Growth Through Collision and Coagulation}

One possible explanation why vortex-induced aggregation is still observed for droplets produced by gas condensation is that these droplets acquire angular momentum through collision and coagulation, see also Ref.~\cite{Escartin2022}. In the following, we present different estimates based on simplified assumptions to explain this effect. These models are naturally of limited accuracy, but they provide some insights, and we hope that they may be a starting point for future theoretical studies.

\begin{figure}[ht]
  \includegraphics[trim=2mm 1mm 9mm 7mm, clip, width=8.6cm]{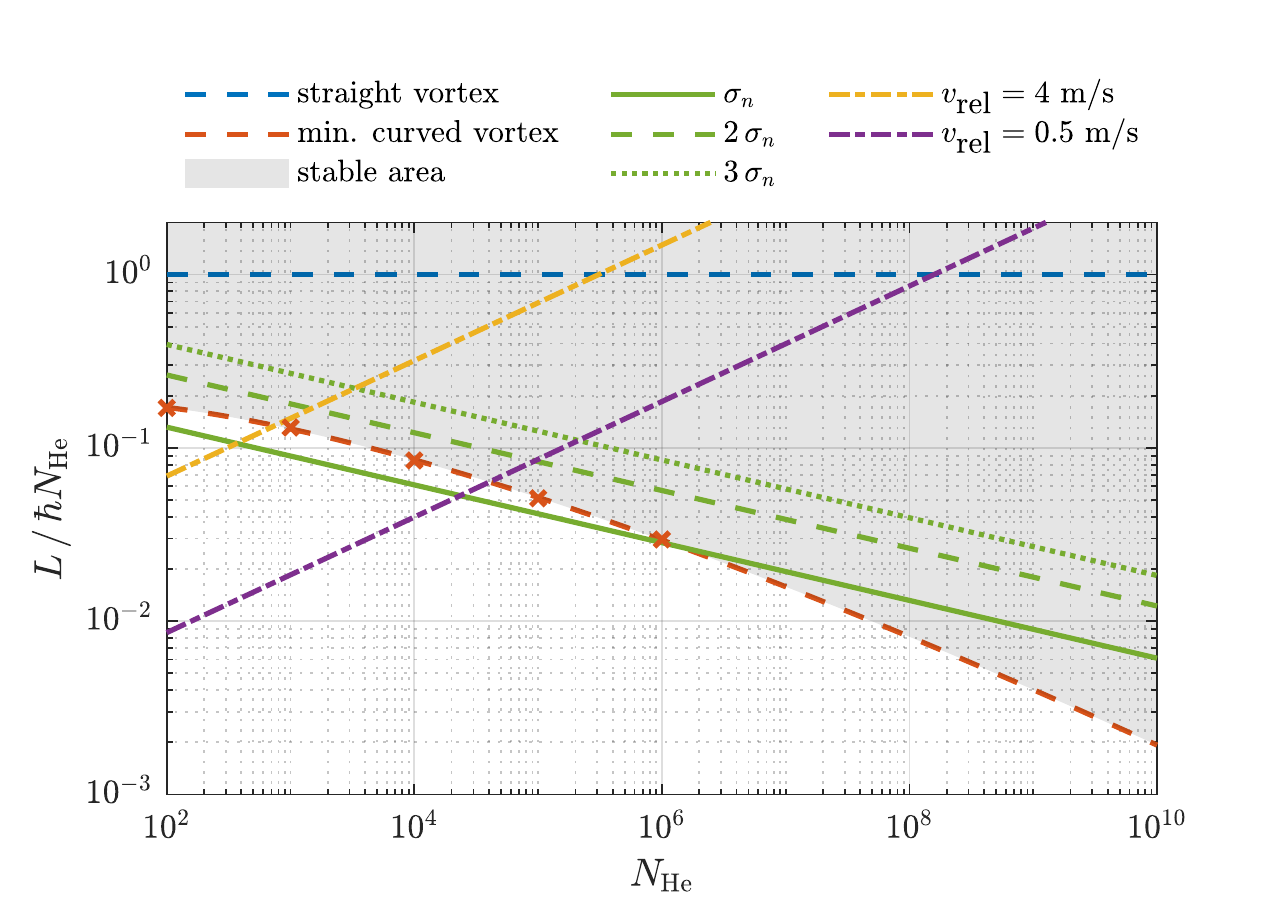}
  \caption{Minimum angular momentum required for a straight (blue dashed line) and curved (red dashed line) vortices for different droplet sizes ${N}_{\mathrm{He}}$. The curved vortex stability curve is extrapolated to larger sizes from the data (red crosses) given in ref.~\cite{Lehmann2003}. The stability region for one vortex is marked with a gray background. Following the model in~\cite{Lehmann2003}, angular momenta $L$ acquired by a single collision of two droplets with ${N}_{\mathrm{He}} / 2$ atoms each, an initial radius $R_\mathrm{i}$ and an average impact parameter of $b = 4 R_\mathrm{i} / 3$ were estimated for different collisional velocities. The dashed dotted lines show $L$ for size independent relative collision velocities of $\SI{0.5}{\metre\per\second}$ (purple) and $\SI{4}{\metre\per\second}$ (yellow). The green lines show $L$ from a collision with a size-dependent Gaussian velocity distribution (centered around $L=0$) at the widths $\sigma_{n}$ (solid), $2 \sigma_{n}$ (dashed) and $3 \sigma_{n}$ (dotted). \label{fig:sm6}}
\end{figure}

Adopting the discussion in Ref.~\cite{Lehmann2003}, the amount of angular momentum acquired during a droplet collision can be approximated by considering two droplets with initial radius $R_\mathrm{i} = r_0 (N_\mathrm{He}/2)^{1/3}$, each containing $N_\mathrm{He} / 2$ atoms, colliding with a relative velocity $v_\mathrm{rel}$ at an average impact parameter $b$. The collisional angular momentum is then given by
\begin{equation}\label{eq:L_col}
    L_\mathrm{col} = \frac{1}{4}\,b\,v_\mathrm{rel}\,m_4\,N_\mathrm{He} 
                   = 2^{-7 / 3}\,\tilde{b}\,v_\mathrm{rel}\,m_4\,r_0\,N_\mathrm{He}^{4/3},
\end{equation}  
where $m_{4}$ is the mass of ${ }^{4} \mathrm{He}$ and $\tilde{b} = b/R_\mathrm{i} = b\,2^{1/3} / r_0 N_\mathrm{He}^{1/3}$ is the dimensionless impact parameter. Two different approaches are considered below to estimate the collision velocity $v_\mathrm{rel}$. The first is based on the speed ratio, defined as $S = \langle v \rangle \sqrt{m / 2 k_\mathrm{B} T_\parallel}$, where $\langle v \rangle \approx \SI{200}{\metre\per\second}$ is the average beam velocity and $T_\parallel$ is the translational temperature parallel to the beam~\cite{Buchenau1990, Harms1997}. The dashed dotted lines in Figure~\ref{fig:sm6} show the angular momentum resulting from droplet collisions at relative velocities of $v_{\mathrm{rel}} = \SI{4}{m.s^{-1}}$ (yellow) and $v_{\mathrm{rel}} = \SI{0.5}{m.s^{-1}}$ (purple), which were calculated from the speed ratios $S \approx \numrange{50}{400}$ reported in Ref.~\cite{Buchenau1990}. From these calculations, a stable vortex should have been generated for all sizes relevant to the present experiment, while more than one vortex is predicted for $>10^8$ atoms. This is not in agreement with our observations. At $N_\mathrm{He}=10^8$, a single curved vortex is mostly observed; a trend which continues for larger droplets. On top of using a $\SI{5}{\micro\metre}$ pinhole nozzle, the maximum speed ratios (minimum relative velocity) reported in Ref.~\cite{Buchenau1990} are limited by the resolution of the instrument and may be considerably underestimated. Furthermore, we like to point out that the clusters in Ref.~\cite{Buchenau1990} are orders of magnitude smaller than the droplets in our experiment and we expect that the relative velocities will decrease with increased size. The speed ratio is expected to increase with nozzle diameter~\cite{Toennies1977} and should be much larger for the conical nozzle used in the present experiment. Hence, an alternative expression is derived.

In the process of creating helium droplets in a supersonic jet, a pre-cooled gas is expanded through a nozzle, causing adiabatic cooling~\cite{Buchenau1990, Hagena1972}. Close to the nozzle throat, where high density prevails, droplets can grow by successive accumulation of single atoms or small clusters. The gas density decreases as it occupies more volume during expansions, and, at some point after expansion, growth by condensation ceases. However, a cluster can grow further by coagulation through a secondary collision process, as is known from other high pressure supersonic jet sources~\cite{Soler1982, Rupp2012}. Here, we assume a simple probabilistic model for growth by successive coagulation to estimate the collision velocities of these clusters. 

The velocity of supersonic jets is typically described by a combination of two Maxwellian distributions with a perpendicular \(v_\perp\) and a parallel \(v_\parallel\) component. The former mainly leads to a reduction of beam flux and is only relevant in the gas condensation region~\cite{Pauly2000}, whereas, the latter remains germane for estimating the angular momentum acquired during droplet growth from collision and coagulation. The parallel velocity component has a Gaussian shape
\begin{align}
    f(v_\parallel) = \sqrt{\frac{m_4}{2\pi k_\mathrm{B} T_\parallel}} \exp\left(-\frac{m_4 v^2}{2 k_\mathrm{B} T_\parallel}\right),
\end{align}
in which \(T_\parallel\) is the temperature parallel to the beam, and, in the inertial frame of the expanding gas, \(v = v_\parallel - \langle v \rangle\), where \(\langle v \rangle\) corresponds to the average beam velocity~\cite{Pauly2000}. Using \(p = m_4 v\) and normalizing the equation above, the momentum probability density function of the atomic gas is given as
\begin{equation}\label{eq:f0distribution}
    f_0(p) = \frac{1}{\sqrt{2\pi}\sigma_0} \exp\left(-\frac{p^2}{2\sigma_0^2} \right), 
\end{equation}
with the standard deviation given by \(\sigma_0 = p_0 = m_4 v_0 = \sqrt{m_4k_\mathrm{B} T_\parallel}\). The velocity spread $v_0\approx\SI{54}{\metre\per\second}$ was estimated assuming that the beam temperature during the coagulation process \(T_\parallel \approx \SI{1.4}{\kelvin}\), which is the geometric mean between the terminal droplet temperature of ($\SI{0.4}{\kelvin}$)~\cite{Hartmann1995, Toennies2004} and the temperature where the isentropes cross the vapor pressure curve ($\approx \SI{5}{\kelvin}$)~\cite{Buchenau1990, Toennies2004}.

Assuming a model where each droplet consisting of $N_{\mathrm{He}}$ atoms is produced from $n$ collisions of equally-sized droplets, i.e., $N_{\mathrm{He}}=2^{n}$, and with an initial momentum distribution given in Eq.~\eqref{eq:f0distribution}, 
the momentum distribution after an arbitrary number of collisions \(n \in \mathbb{N}\) may be estimated from the convolution of the momentum distribution of  \(n-1\) collisions with itself
\begin{align}
    f_n(p) &= (f_{n-1} * f_{n-1}) (p)
    = \frac{1}{\sqrt{4\pi}\sigma_{n-1}}  \exp\left(-\frac{p^2}{4\sigma_{n-1}^2}\right) \nonumber\\
    &= \frac{1}{\sqrt{2\pi}\sigma_n}  \exp\left(-\frac{p^2}{2\sigma_n^2}\right),
\end{align}
from which follows that $\sigma_n = \sqrt{2} \sigma_{n-1} = \sqrt{2^n} \sigma_{0}$. This model rests on the fact that large clusters/droplets generated in the gas phase regime grow largely by coagulation~\cite{Soler1982, Ratner1999, Jansen2011}. The width of the distributions after $n$ collisions is
\begin{equation}
  \sigma_{n}=2^{n / 2} \sigma_{0}=\sqrt{N_{\mathrm{He}}} \cdot m_{4} v_{0}=p_{n}=N_{\mathrm{He}} \cdot m_{4} v_{n}.
\end{equation}
This equation implies that collisional velocities decrease upon successive collisions, which is equivalent to a size-independent translational temperature $T_\parallel$ and to the mass-dependent relative velocity $v (N_\mathrm{He}) = \sqrt{2 k_\mathrm{B} T_\parallel/N_\mathrm{He} m_4}$. Within a translational freezing model for an average droplet size estimation, Ref.~\cite{Buchenau1990} also assumed a size-independent translational temperature. Considering the standard deviation $\sigma_0$ as an estimate of the relative velocity between two droplets containing \(N_\mathrm{He} / 2\) atoms each,
\begin{equation}
    v_\mathrm{rel} \equiv v_{n-1} = v_0\, \left(N_\mathrm{He} / 2 \right)^{-1/2}, \label{eq:sizeDependentCollisionVelocity}
\end{equation}
which yields collision velocities of \(v_\mathrm{rel} = \SI{7.8e-2}{\metre\per\second}\), \(\SI{7.8e-3}{\metre\per\second}\), and \(\SI{7.8e-4}{\metre\per\second}\) for \(N_\mathrm{He} = 10^6\), \(10^8\) and \(10^{10}\), respectively. While the assumption that equally-sized droplets coagulate is certainly a simplification, we expect Eq.~\eqref{eq:sizeDependentCollisionVelocity} to remain applicable even for the coagulation of differently-sized droplets since the size-dependent velocity originates from momentum conservation. These estimated collision velocities are much smaller than those estimated from the measured speed ratios discussed above~\footnote{The model is certainly based on simplifications and assumptions, the details of growth in the initial phase, the role of evaporation due to the heat of condensation and when the transition from growth by monomer addition to coagulation occurs are not explicitly considered. However, we would like to point out that the translational temperature of clusters grown out of the gas phase can be estimated from Ref.~\cite{Buchenau1990} (peak 2) and using the known scaling laws for cluster sizes~\cite{Hagena1972, Hagena1992, Dorchies2003} to the very low values in the range of \SI{\sim 0.1}{\kelvin}. These low values seem reasonable because the clusters are only very loosely bound at the onset of condensation, and very small relative velocities and translational temperatures are required to generate them, thus suggesting that the relative collisional velocities tend to be even smaller estimated in our model above.}. Substituting Eq.~\eqref{eq:sizeDependentCollisionVelocity}, as obtained from our model, into Eq.~\eqref{eq:L_col}, the angular momentum expression as a result of droplet collisions is approximately given as
\begin{equation}\label{eq:LFromCollision}
    L = 2^{-11/6} \, \tilde{b} \, m_4 \, r_0 \, v_0 \, N_\mathrm{He}^{5/6}. 
\end{equation}
The results of this estimation are shown for $\sigma_{n}, 2 \sigma_{n}$, and $3 \sigma_{n}$ in Fig.~\ref{fig:sm6} in solid, dashed and dotted dashed green, respectively. These values reflect the width of the velocity distribution, which is centered around zero.

From these considerations, it is very unlikely to create a straight vortex from a single collision at low relative velocities. At the same time, the probability for creating a stable single curved vortex increases with droplet size, predicting that more than $50\%$ of collisions induce sufficient angular momentum for $N_\mathrm{He} \geq \num{5e7}$, which is in reasonable agreement with our observation. It must be noted that the extrapolation of curved vortex stability to larger droplets, as done in Fig.~\ref{fig:sm6}, must be taken with caution. Only ripplon excitations were considered in Ref.~\cite{Lehmann2003}; a consideration that suffices for droplets with $N_{\mathrm{He}} \leq 10^{6}$ because phonons are not thermally excited at $\SI{0.4}{K}$ for those sizes. For the droplet sizes presented here, phonons might already play an essential role in the storage of angular momentum and could bend the lower bound of a stable curved vortex upward to higher $L$ values (red dashed line in Fig.~\ref{fig:sm6}), especially considering that phonon emission is a dissipative channel for the decay of vortices at low temperatures ~\cite{Paolett2011, Harris2016, Madeira2020}. Another aspect that has not been discussed up to now is that coagulation will heat up the droplet, followed by evaporative cooling. This process reduces the droplet size and may also reduce the angular momentum of the droplet~\cite{Lehmann2004}. Nonetheless, Fig.~\ref{fig:sm6} can give us an idea about vortex stability conditions, especially for $N_{\mathrm{He}} \leq 10^{8}$, where the majority of excitations are ripplons, see Fig.~8.1 in Ref.~\cite{Tanyag2018}.

\section{Generation of Helium Droplets Using Different Nozzles}

The nozzle stagnation conditions define the thermodynamic states (enthalpy and entropy) of the expanding fluid. For the generation of superfluid helium nanodroplets, different regimes are identified depending on which side of the saturation vapor curve the fluid would cross in an adiabatic isentropic expansion~\cite{Buchenau1990, Harms1997, Grisenti2003, Toennies2004, Slenczka2022}. Stagnation conditions crossing from the gas phase are grouped under the gas condensation regime, while those crossing from the liquid phase are under liquid fragmentation.   

\begin{figure}[ht]
  \includegraphics[trim=4mm 7mm 8.5mm 6.25mm, clip, width=.48\textwidth]{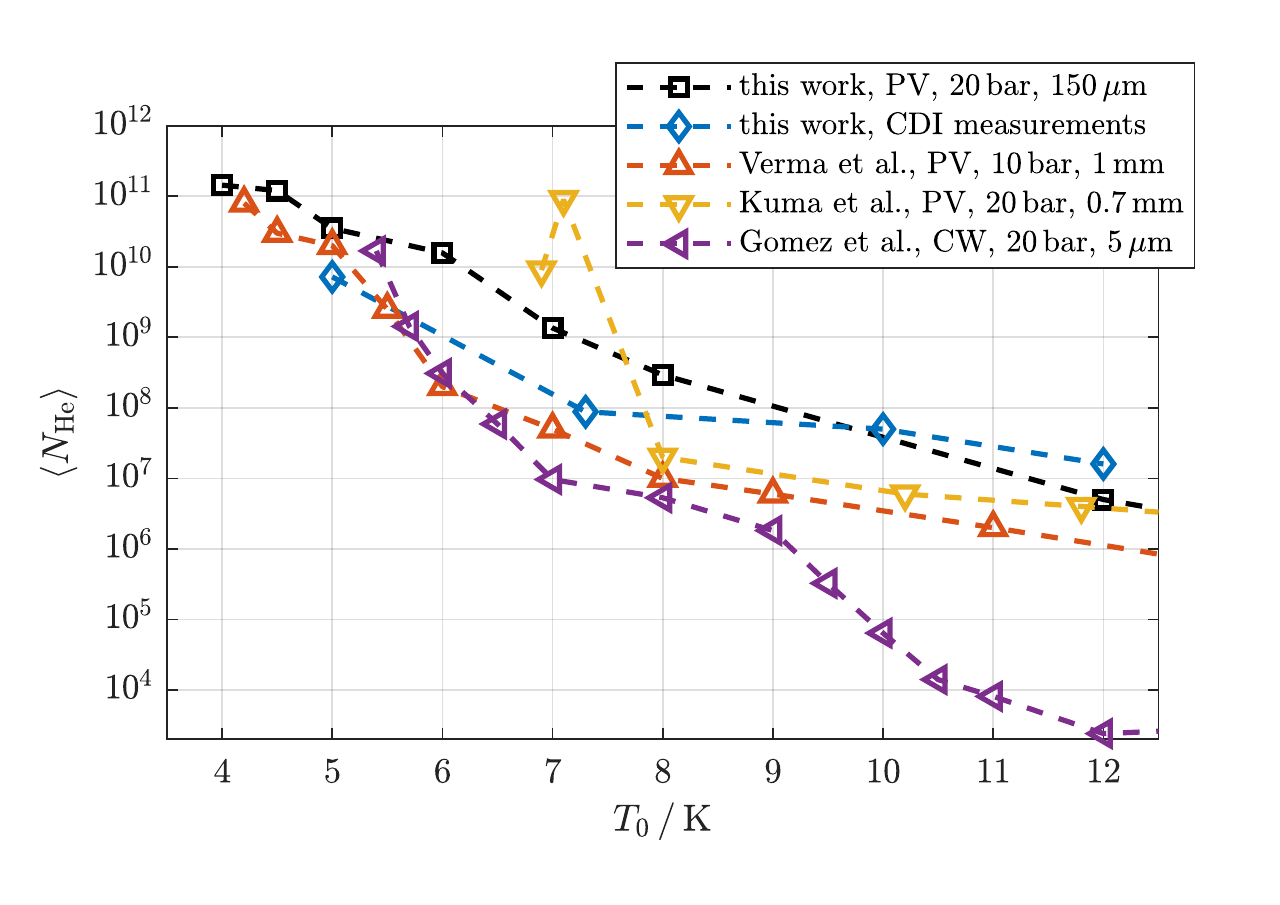}
  \caption{A comparison of average droplet sizes measured using different droplet sources. The purple \rotatebox[origin=c]{90}{$\triangle$} were obtained from Ref.~\cite{Gomez2011}, which used a $\SI{5}{\micro\metre}$ pinhole nozzle with continuous flow (CW). The yellow \rotatebox[origin=c]{180}{$\triangle$} and the red \rotatebox[origin=c]{0}{$\triangle$} are from Ref.~\cite{Kuma2017} and  Ref.~\cite{Verma2018}, respectively; both employed a Parker valve (PV) and a conical nozzle with an opening angle of $\ang{90}$ but with a different throat diameter. The black $\square$ are from Ref.~\cite{Behrens2019MA}, which used the same droplet source as in this work but a different detector. The droplet sizes for these four data sets were measured employing the "titration" technique from Ref.~\cite{Gomez2011}. Finally, the blue $\lozenge$ are the sizes determined from diffraction images from this experiment. Except for the measurements from Ref.~\cite{Verma2018}, the stagnation pressures were constant at $\SI{20}{bar}$. The dashed lines are meant as a guide to the eye.\label{fig:sm7}}
\end{figure}

In many molecular beam setups for the generation of helium droplets, the use of a $\SI{5}{\micro\metre}$ pinhole nozzle is standard, and its operation and performance are known~\cite{Buchenau1990, Toennies2004, Gomez2011, Slenczka2022}; for a recent schematic of this pinhole nozzle, see Fig. S1 in the supplementary material of Ref.~\cite{Tanyag2020}. Figure~\ref{fig:sm7} shows the average droplet sizes from a pinhole nozzle measured employing the "titration" technique described in Ref.~\cite{Gomez2011}. Small droplets containing roughly $10^4$ atoms are generated from gas condensation ($T_0 \geq \SI{10}{\kelvin}$), while larger droplets with more than $10^6$ atoms are produced from the liquid fragmentation ($T_0 < \SI{5}{\kelvin}$) regime~\cite{Toennies2004, Slenczka2022, Gomez2011}. As revealed by x-ray coherent diffractive imaging experiments, a significant fraction ($>0.30$) of these larger droplets have $AR$ greater than $1.05$~\cite{Bernando2017, Tanyag2018b} and are likely studded with quantum vortices~\cite{Gomez2014, OConnell2020}. Therefore, experiments wanting to image vortex-free droplets and the nanostructures assembled in them must obviate the use of pinhole nozzles while keeping in mind the droplet size range that still permits size determination from a diffraction image. Current limitations to x-ray imaging geometry constraint this range within $N_\mathrm{He} = 10^6$ up to $10^{11}$~\cite{Tanyag2022}.

Droplets generated from the gas condensation regime are likely vortex-free since their creation involves many statistical processes from the expanding cold helium gas. As such, they are less likely to acquire angular momentum from the shear forces introduced by the co-flowing helium, which is in contrast to those generated from the liquid fragmentation regime. Nonetheless, nucleation of quantum vortices remains feasible from droplet collision and coagulation, see the discussion in the preceding section and those in Refs.~\cite{Lehmann2003, Escartin2022}.

\begin{figure*}
  \includegraphics[width=\textwidth]{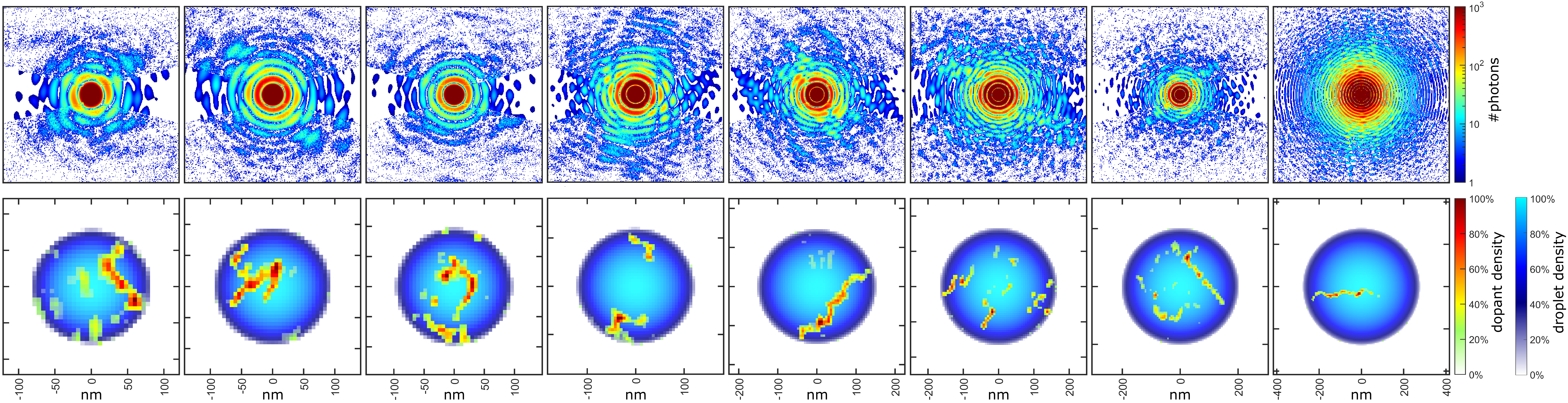}
  \caption{Vortex structures at a higher nozzle stagnation pressure of $p_{0}=\SI{60}{bar}$ and temperature of $T_{0}=\SI{12}{K}$. The top row displays measured diffraction intensities, where the masked regions are completed by the reconstructed intensity values. The bottom row shows the respective reconstructed dopant structures. Vortices are much more prominent in this data set compared to data recorded at a lower stagnation pressure of $p_{0}=\SI{20}{bar}$. \label{fig:sm8}}
\end{figure*}

Following known scaling laws, particularly for rare-gas clusters~\cite{Hagena1972, Hagena1981, Wormer1991, Karnbach1993, Dorchies2003}, large clusters or droplets can be generated from the gas condensation regime by enlarging the nozzle diameter or by shaping the nozzle channel (conical, trumpet, de Laval) such that the outflowing fluid is confined to a small angle with respect to the axial flow direction, concentrating the on-axis beam intensity. Ref.~\cite{Even2014} presents a Direct Simulation Monte Carlo (DSMC) calculation of axisymmetric flow at a low stagnation pressure of $\SI{0.3}{bar}$ for three types of nozzles: sonic, which is similar to the pinhole nozzle; conical-shaped; and trumpet-shaped. Shaping the flow effectively slows down the expansion process and increases the rate of collisions that promotes cluster growth through condensation and coagulation~\cite{Hagena1972, Hagena1981, Even2014}; this effect is similar to the "ripening" process described in Ref.~\cite{Verma2018}. In contrast, expansion from a smaller pinhole nozzle is faster, which stunts cluster growth~\cite{Hagena1972, Even2014}; hence, only smaller clusters are created. This effect is demonstrated in Fig.~\ref{fig:sm7} where the measured droplets sizes in the gas condensation regime vary by two to three orders of magnitudes between the pinhole and conical nozzles. We note that the measured average droplet of $N_\mathrm{He} = 10^8$, generated using our conical nozzle at stagnation conditions of 20\,bar and 10\,K, is in accordance with that predicted by scaling laws used in cluster generation, where large clusters are generated through coagulation from the expansion of cold gas. 

Since the equivalent nozzle diameter for shaped-nozzles can be several hundreds of times larger than pinhole nozzles~\cite{Hagena1972, Hagena1981, Knuth1997}, they are operated in tandem with pulse valves to prevent inundating the capacity of turbomolecular pumps and to maintain low background pressure. Ideally, pulsed nozzles should reach continuous flow conditions at longer valve opening times~\cite{Morse1996, Christen2013, Even2014}, as was done in this experiment. In many cases, however, the opening and closing of the valve is affected by the phase of outflowing fluid (which for some stagnation conditions may be a combination of gas and liquid) that could lead to complicated droplet beam profile~\cite{Even2014}. To date, pulsed conical-~\cite{Slipchenko2002, Yang2005, Kuma2017, Verma2018} and trumpet-shaped~\cite{Pentlehner2009, Katzy2016, Pandey2021} nozzles have been used for generating larger helium droplets in the gas condensation regime; both with increasing use in a variety of contemporary experiments. 

The trumpet-shaped nozzle attached to an Even-Lavie valve~\cite{Pentlehner2009, Even2014} has already been used in earlier coherent diffractive imaging experiments~\cite{Rupp2017, Langbehn2018}, see Refs.~\cite{Pentlehner2009, Even2014} for the schematic of the Even-Lavie valve with a trumpet-shaped nozzle. Note that the stagnation condition used in operating the Even-Lavie valve for these imaging experiments follows an isentropic expansion trajectory that crosses the saturation vapor curve from the liquid phase; $\SI{5.4}{K}$, $\SI{80}{bar}$ is isentropic to $\SI{\sim 4}{K}$, $\SI{20}{bar}$~\cite{McCarty1973}. High stagnation pressures (or high flow velocities) are susceptible to turbulence, and the droplets are likely to acquire vorticity from the turbulent flow, enough to nucleate multiple vortices and distort the droplet shape. A similar effect was observed for the conical nozzle used in this experiment. For instance, no vortex structures were be observed at $T_{0}=\SI{12}{K}$ and $p_{0}=\SI{20}{bar}$, but increasing $p_{0}$ to $\SI{60}{bar}$ at the same $T_{0}$, xenon-traced vortices become prominent in the data set, see examples shown in Fig.~\ref{fig:sm8}.

Overall, the shape of the nozzle is an additional parameter for controlling the condensation process, which in turn facilitates the generation of large droplets with $> 10^6$ atoms from the gas condensation regime. Additionally, since these large droplets are likely formed stochastically through collision with other helium droplets at some distance away from the nozzle that may preclude the acquisition of angular momentum from shear forces induced by the co-flowing helium fluid, these large droplets acquire less angular momentum. Nevertheless, in order to ascertain which type of nozzle favors the creation of large vortex-free droplets, further imaging experiments using different kinds of nozzles are needed.

\bibliography{library}